\documentclass[a4paper]{aa}

\usepackage{graphicx,float,psfrag}
\usepackage{tabularx}
\usepackage{latexsym,amsmath,amssymb}
\usepackage{natbib}
\usepackage{verbatim}
\usepackage{rjwmath}
\usepackage{multirow}

\bibpunct{(}{)}{;}{a}{}{,}

\def \george    {G. Younes}
\def \delphine  {D. Porquet}
\def \bassem    {B. Sabra}
\def \nicolas   {N. Grosso}
\def \jreeves   {J.~N. Reeves}

\def \strasbourg {Observatoire Astronomique de Strasbourg, Universit\'e de Strasbourg, CNRS, UMR7550, 11 rue de
l'Universit\'e, F-67000 Strasbourg, France}

\def \ndu {Department of Physics \&\ Astronomy,  Notre Dame University-Louaize, P.O.Box 72, Zouk Mikael, Lebanon}
\def \keele {Astrophysics Group, School of Physical \&\ Geographical Sciences, Keele University, Keele, Staffordshire ST5 5BG}

\def \maryland {Department of Physics, University of Maryland Baltimore County, Baltimore, MD 21250, USA} 

\newcommand {\hst} {\textrm{HST}}
\newcommand {\xmm} {\textsl{XMM-Newton}}
\newcommand {\chandra} {\textsl{Chandra}}

\def \rsun {\ifmmode$R$_{\odot}\else R$_{\odot}$}

\def \hcm {\hbox {\ifmmode $ atoms cm$^{-2}\else atoms cm$^{-2}$\fi}}

\def\approxgt{\mathrel{\hbox{\rlap{\lower.55ex \hbox {$\sim$}}
        \kern-.3em \raise.4ex \hbox{$>$}}}}
\def\approxlt{\mathrel{\hbox{\rlap{\lower.55ex \hbox {$\sim$}}
        \kern-.3em \raise.4ex \hbox{$<$}}}}

\def\arcsec{\hbox{$^{\prime\prime}$}}

\def \eddratio {$L_{\rm 2-10~keV}/L_{\rm Edd}$}
\def \alphaox  {$\alpha_{\rm ox}$}

\begin{document}

\title{Study    of     LINER    sources    with     broad    H$\alpha$
  emission.\\   Spectral  energy   distribution   and  multiwavelength
  correlations}

\author{\george\inst{1} \and \delphine\inst{1} \and \bassem\inst{2} \and \jreeves\inst{3,4} \and \nicolas\inst{1}}

\institute{\strasbourg \and \ndu \and \keele \and \maryland}

\date{Received / Accepted}

\authorrunning{G. Younes et al.}

\titlerunning{Multiwavelength properties of LINER~1s}

\abstract{The geometry and physical  properties of the accretion mode,
  and the radiative processes  occurring in AGN-powered low ionization
  nuclear  emission-line  regions (LINERs)  remain  a  riddle. Both  a
  standard     thin     accretion     disk    and     an     inner-hot
  radiatively-inefficient    accretion   flow    (RIAF)    have   been
  invoked. Models  depending on only a  jet have also  been invoked to
  explain  the  broad-band   spectral  energy  distribution  (SED)  of
  LINERs.}{We attempt  to infer the accretion  mechanism and radiative
  processes   giving   rise   to    the   SEDs   of   a   well-defined
  optically-selected sample of LINERs  showing a definite detection of
  broad  H$\alpha$  emission (LINER~1s).}{We  construct  SEDs for  six
  LINER~1s with {\it simultaneous UV  and X-ray fluxes}, and we looked
  for   multiwavelength,   radio   to   X-ray   and   UV   to   X-ray,
  correlations.}{At a  given X-ray luminosity, the average  SED of the
  six  LINER~1s in  our sample:  (1) resembles  the SED  of radio-loud
  quasars  in  the  radio  band,  $<\log~R_{\rm  X}>\approx-2.7$,  (2)
  exhibits  a weak UV  bump, $<$\alphaox$>\approx-1.17\pm0.02$  with a
  dispersion $\sigma=0.01$, and (3)  displays a X-ray spectrum similar
  to radio-quiet  quasars.  The bolometric  luminosities inferred from
  the SEDs of these LINER~1s  are extremely faint, at least two orders
  of  magnitude  lower than  AGN.   The  X-ray bolometric  correction,
  $\kappa_{\rm 2-10~keV}$, of our sample  is lower than in the case of
  AGN,  with a mean  value of  16.  We  find a  strong anticorrelation
  between the radio loudness parameter, $R_{\rm X}$, and the Eddington
  ratio  for our  sample, confirming  previous results.   Moreover, we
  find  a positive correlation  between the  radio luminosity  and the
  X-ray luminosity  which places AGN-powered LINERs,  on a radio-power
  scale,  right between  low  luminosity Seyferts  and low  luminosity
  radio  galaxies.   We  complement  our \alphaox\  list  with  values
  derived  on  a  well  defined  sample  of  UV-variable  LINERs,  and
  establish     a      strong     positive     correlation     between
  \alphaox\ (considering negative values)  and the Eddington ratio, in
  contrast  to the  correlation found  for luminous  AGN.   Lastly, we
  tested two  different fundamental planes existing  in the literature
  on  our sample,  in an  attempt to  put constraints  on  the debated
  origin of the X-ray emission, ``RIAF versus jet''.  The results came
  contradictory  with  one  pointing  toward  a  RIAF-dominated  X-ray
  emission process and the other pointing toward a jet domination.}{}


\maketitle

\section{Introduction}
\label{sec:intro}

Our  nearby  universe is  mostly  populated  by  galaxies showing  low
nuclear  activity \citep[see][and  references therein]{ho08aa:review}.
The  most  common component  of  these  nearby  low luminosity  active
galactic nuclei (LLAGN\footnote{Hereinafter, the term LLAGN designate
  the definite existence of an AGN  at a given galaxy center, but with
  an  Eddington  ratio  $\le10^{-3}$.})   is  low  ionization  nuclear
emission-line regions, LINERs  \citep{heckman80aap}.  These LINERs are
characterized  by  optical  spectra  dominated by  neutral  or  singly
ionized  species  (e.g.,  O~I),   hence,  non-AGN  continua  could  in
principle give rise to their  optical spectra.  Indeed, since the time
of their establishment as a class, the excitation mechanism of LINERs
has been a matter of an ongoing debate and could be explained in terms
of: shock heated gas \citep{dopita95apj:shocliner}, starburst activity
\citep{                                terlevich95mnras:starburstliner,
  alonso-herrero00ApJ:starburstinliners},    or    a   LLAGN.     Many
multiwavelength studies were attributed to this subject, looking for a
radio, sometimes  variable, core \citep{4278nagar05aap}  or a variable
UV  core \citep{maoz05apj:linervarUV}.   Nevertheless,  the most  used
tool to search for an active nucleus  in a LINER is to look for a hard
2-10~keV unresolved  core that  could not be  due to  diffuse emission
from   shock  heated  gas   or  from   unusually  hot   stars  \citep{
  terashima00ApJ:liners,             ho01apjl,dudik05apj:hardcoreliner,
  flohic06apj,         gonzalezmartin06aa,         gonzalezmartin09aa,
  zhang09apj:llagnxray}.   How do  LINERs harboring  a  low luminosity
active nucleus compare to luminous Seyfert galaxies and quasars?

Active galactic nuclei (AGN), including Seyfert galaxies and quasars,
emit a  broad band spectral  energy distribution (SED) from  radio all
the way  up to X-rays,  sometimes even gamma-rays. The  most prominent
feature in the  SED of AGN is  a broad UV excess, known  as the ``big
blue  bump''  \citep[e.g.,][]{elvis94apjs:quasar}.   This  UV  excess,
along with  many other signatures characterizing AGN;  e.g., the high
bolometric luminosities ($L_{\rm bol}\ge10^{44}$) and Eddington ratios
($L_{\rm bol}/L_{\rm  Edd}\ge10^{-2}$, where $L_{\rm  Edd}$ represents
the    Eddington     luminosity;    \citealt{    porquet04aa:pgquasar,
  vasudevan09MNRAS:agnSED,vasudevan09MNRAS:agnSED2,grupe10ApJS:agnSED}),
the  rapid   X-ray  variability  on  time-scales  of   hours  to  days
\citep{turner99apj:seyvar},  and  the  gravitationally  redshifted  Fe
emission line profile \citep{nandra07mnras:felinesey}, are believed to
be achieved  through accretion onto a supermassive  black hole (SMBH),
in the  form of  a geometrically thin  optically thick  accretion disk
\citep{shakura73aa, shields78Natur:UVbump, Malkan82ApJ:softexce}.

Contrary  to  classical AGN,  where  at  least  a somewhat  universal
accretion  mode  is  thought  to  occur,  the  accretion  physics  and
radiative  processes  taking place  in  LLAGN, including  AGN-powered
LINERs, have been a riddle so far.  LINERs are extremely faint sources
with X-ray luminosities almost never surpassing $10^{43}$~erg~s$^{-1}$
and Eddington  ratios lower  than $\sim10^{-3}$, and  explaining their
faintness is a very challenging task.

The  geometrically thin accretion  disk, believed  to power  AGN, has
been  invoked  as   a  LINER  source  of  power   as  well.   From  an
observational point  of view, \citet{maoz07MNRAS},  using high angular
resolution  multiwavelength  observations   of  13  UV-variable  LINER
sources,  demonstrated   that  the  luminosity   ratios  in  different
wavebands, mainly UV  to X-ray and radio to UV,  follow the same trend
as  luminous Seyfert  galaxies. The  authors  did not  find any  sharp
change  in the  spectral  energy distribution  (SED)  of their  sample
compared to more luminous Seyfert and quasar nuclei, suggesting that a
thin   accretion   disk  may   persist   at   low  Eddington   ratios.
\citet{pianmnras10}  combined  {\sl   Swift}  XRT  X-ray  fluxes  with
simultaneous  UV fluxes coming  from the  UVOT instrument  and showed,
similar to \citet{maoz07MNRAS}, that the  SED and the UV to X-ray flux
ratios of their  sample of 4 LINERs are consistent  with those of more
luminous  sources,  hence,  implying  that  LINERs  may  have  similar
accretion and radiative processes at their center compared to luminous
Seyfert nuclei.

On  the  other  hand,  the  faintness of  LINER  sources  compared  to
classical AGN has been  attributed to a different accretion mechanism
owing to some  observational contrast between the two  classes. In the
X-ray  domain,  no broad  nor  narrow  Fe~K$\alpha$  emission line  at
6.4~keV have  been detected in the  spectra of the  LINER sources with
the    highest     signal    to    noise     ratio    \citep[][Paper~1
  hereinafter]{ptak04apj:ngc3998, younes11AA:liner1sXray}, with only a
few   sources   exhibiting    short   time-scale   X-ray   variability
\citep{ptak98apj:variance,  awaki01pasj:varliner, binder09apj:ngc3226,
  younes10aa:ngc4278}.   Furthermore,   early  investigations  of  the
multiwavelength properties of LLAGN, including LINERs, indicated that
their  SED might  depart  from  the standard  SED  of classical  AGN,
especially in the UV band. \citet{ho96ApJ:M81SED} reported the absence
of  the  UV  bump  in  the  SED  of the  LINER  source  M~81,  so  did
\citet{nicholson98MNRAS:sed}  for the LINER  nucleus of  NGC~4594. The
evidence for  the faintness of the  UV component in the  SED of LINERs
started to  pile up  with the study  of bigger  samples.  \citet[][see
  also   \citealt{   eracleous10:linersed}]{ho99sed}  collected   high
spatial  resolution  multiwavelength  fluxes  for a  sample  of  seven
LLAGN.   The author  stated that  the SED  of the  full  sample looks
markedly different than  the SED of luminous AGN,  noting that the UV
emission  is  weaker,  lacking   the  canonical  UV  bump.   Moreover,
\citet{ho99sed}  signalized  that all  of  his  LLAGN  SEDs could  be
considered as radio-loud sources.  These observational dissimilarities
between   LINERs/LLAGN   and   classical   AGN  may   indicate   the
truncation/disappearance   of   the  thin   accretion   disk  at   low
luminosities, and therefore,  a different accretion mechanism powering
the emission in AGN-powered LINERs. Indeed, it has been suggested that
when   the  mass  accretion   rate  falls   below  a   critical  value
$\dot{M}_{crit}$,   for   which   LINERs/LLAGN  clearly   belong   to
\citep{ho09apj:riaf}, the density of the disk could become too low for
radiative cooling to  be effective.  The trapped heat  will expand the
thin accretion disk  into a pressure-supported thick disk  with a very
low        radiative         efficiency        \citep[see        ][for
  reviews]{quataert01aspc:riaf, yuan07ASPC:adaf,narayan08:riafreview}.
Such radiatively inefficient accretion flow (RIAF) models successfully
explained the spectral energy distribution  of a large number of LINER
sources    \citep{    quataert99apj:m81ngc4579,   gammie99apj:ngc4258,
  ptak04apj:ngc3998,         nemmen06apj:ngc1097,        wu07apj:riaf,
  yu11ApJ:maozsed}.

The early RIAF  solutions showed that the gravitational  energy of the
accretion flow  could be lost,  not only through convection,  but also
through outflowing winds  and jets \citep[e.g.,][]{ narayan94ApJ:adaf,
  narayan95ApJ:adaf,     blandford99MNRAS:jet}.      Numerical     and
magnetohydrodynamic  (MHD)  simulations  confirmed  this  finding  and
showed that RIAFs are associated with strong outflows \citep[e.g.,][]{
  igumenshchev00ApJS:adaf, hawley02ApJ:adaf, igumenshchev04PThPS}, and
even a component along the axis of the accretion flow in the form of a
collimated     relativistic    jet    \citep{mckinney06MNRAS:adafjet}.
Observational evidence strengthened the  idea of low accreting objects
to  produce   outflows  and  relativistic  jets   since  an  important
percentage  of LINERs/LLAGN showed  arcsecond or  milliarcsecond jets
\citep[e.g.,][]{nagar01:radobs}. Even  the sources lacking  a jet-like
component  on   milliarcsecond  scale  were  observed   to  harbor  an
unresolved radio core that  have a high surface brightness temperature
and a flat or slightly  inverted radio spectrum, classic signatures of
the  presence of  a relativistic  jet  \citep{ blandford79ApJ:jetagn}.
Through  synchrotron and synchrotron  self-Compton processes,  the jet
could  contribute to the  broad-band emission  of these  low accreting
sources,  sometimes  even  dominating  the  accretion  flow  emission.
Indeed, what is giving rise to the broad-band spectrum, especially the
hard X-ray emission, is a  highly debated subject, with some crediting
completely  the  RIAF models,  others  leaning  towards  a jet  origin
\citep{falcke00A&A:jetmodSgrA, yuan02A&A:jetmodSgrA, yuan02A&A:jetmod,
  wu07apj:riaf, markoff08ApJ:m81}.

The  previous  SED  studies  of  LINERs/LLAGN,  however,  dealt  with
non-simultaneous  UV and  X-ray observations,  sometimes  separated by
several  years.   LINERs,  similar  to  AGN, show  a  high  level  of
variability on  years time-scales,  mainly in the  UV and  X-ray bands
\citep[Paper1;][]{maoz05apj:linervarUV,       pianmnras10}.       This
variability  factor if ignored  could lead  to inconsistencies  in the
LINER  SEDs,  strengthening or  undermining  the  UV  excess in  these
particular sources.   Since the accretion physics rely  heavily on the
strength of  the UV  excess relative to  X-rays, this  important point
should continuously be kept in mind.

Finally,  there is an  increasing number  of evidence  indicating that
LINERs/LLAGN exist in a different spectral state compared to classical
AGN, displaying a behavior  similar to transient X-ray binaries (XRBs,
\citealt{done02RSPTA:XRBs,                   remillard06ARA&A:xrbState,
  done07A&ARv:XRBs}).  In  this context LINERs/LLAGN  would be similar
to  XRBs during  their  low/hard state,  showing  very low  bolometric
luminosities (e.g., Paper~1), and  classical AGN would be analogous to
high/soft state XRBs,  with Eddington ratios $\ge0.1$ \citep[e.g.,][]{
  sobolewska09MNRAS:AGNstate}.             Recently,           \citet{
  sobolewska11MNRAS:aox}  predicted, by  simulating the  spectra  of a
sample  of AGN  based on  the evolution  pattern of  the  transient BH
binary GRO  J1655--40, that  the UV-to-X-ray flux-ratio  dependence on
the  Eddington  ratio  should  change  signs  at  $L_{\rm  bol}/L_{\rm
  Edd}\approx10^{-2}$.   This  change  of  sign,  if  observed,  could
represent another  evidence supporting the  idea of LINERs being  in a
different spectral state compared to classical AGN.

In this work, we revisit the  SEDs of AGN-powered LINERs in an attempt
to put  constraints on the  accretion mode and radiative  processes of
this class.   To ensure the  presence of an  AGN at the center  of our
LINERs, only those  showing a definite detection of  a broad H$\alpha$
emission   where   selected   \citep[Paper1;][LINER~1s   hereinafter]{
  ho97apjs:broadHal}.   We briefly  re-introduce our  sample  in \S~2,
\S~3 represents  the data compilation used to  construct reliable SEDs
with simultaneous  UV and X-ray fluxes,  and gives the  results of the
different  flux  ratios.  We  give  the  properties  of the  different
LINER~1 SEDs  and of their mean,  which represents to  some extent the
mean SED of  all AGN-powered LINERs, in \S~4.   We discuss our results
in \S~5 in  the context of accretion physics  and radiative properties
occurring in  AGN-powered LINERs.  Finally, we  summarize our findings
in \S~6.  In the remainder  of this paper, luminosities are calculated
using the distances given in  Table~1 of Paper~1 derived with a Hubble
constant $H_{0}=75$~km~s$^{-1}$~Mpc$^{-1}$.

\section{The sample}
\label{sec:sample}

The LINER sample studied here has been introduced in a companion paper
\citep[][Paper~1   hereinafter]{younes11AA:liner1sXray}.   The  sample
comprises only  the LINERs showing  the definite detection of  a broad
H$\alpha$    emission    \citep[LINER~1s,][]{ho97apjs:broadHal},   and
observed with either the  \chandra\ and/or the \xmm\ telescopes.  This
resulted in 13  LINER~1s. This class of LINER~1s  ensures the definite
existence of an AGN at the  center of all of the selected galaxies and
its responsibility for the excitation of the detected optical emission
lines  \citep{terashima00ApJ:liners}.  Moreover  and according  to the
unification scheme  of AGN, the  central emission of  LINER~1s should
not  be  affected   by  the  circumnuclear  dust  in   the  torus  (if
existing). The reader  is referred to Paper~1 for  more details on the
general properties of the sample.

\section{Results}

\subsection{SED construction with simultaneous UV and X-ray fluxes}
\label{reliable-SEDs}

The nuclear flux  of LINERs could easily be  contaminated by the light
of the underlying host  galaxy.  Therefore, high angular resolution at
all wavelengths is essential.  Another complication one should keep in
mind  is that  LINERs  could be  highly  variable on  months to  years
timescales,    especially    in    the    UV    and    X-ray    domain
\citep[Paper~1;][]{maoz05apj:linervarUV,  pianmnras10}.  Fluxes coming
from simultaneous observations  are therefore extremely crucial.  With
these two points in mind, we  detail in this section the tools we used
to construct the SED of the sources in our sample of LINER~1s with the
most reliable  multiwavelength data, having simultaneous  UV and X-ray
fluxes.

To rule  out the variability factor,  mostly noticeable in  the UV and
X-ray  bands,  we used  UV  observations  coming  from the  optical/UV
monitor (OM)  telescope, onboard \xmm,  taken with either the  UVM2 or
the   UVW2   filters   (effective   wavelengths   of   2310~\AA\   and
2120~\AA\  respectively). These  UV fluxes  are simultaneous  with the
\xmm\ X-ray EPIC fluxes. Despite  the fact that the angular resolution
of the  OM is much lower than  the \hst\ one, LINER  sources appear as
point-like  sources in the  UV bands,  with the  emission of  the host
galaxy left undetected. Six sources and a total of 7 observations were
observed  with  the  OM with  either  the  UVM2  or the  UVW2  filters
(NGC~315,  obs.   ID:  0305290201;  NGC~3226,  obs.   ID:  0101040301;
NGC~3718,  obs.  ID:  0200430501 and  0200431301; NGC~3998,  obs.  ID:
0090020101;   NGC~4143,   obs.    ID:   0150010601;   NGC~4278,   obs.
ID:0205010101, see Paper~1 for  details on the observation logs), with
NGC~315 being the  only source observed with both  filters.  We derive
count rates,  using the {\sl omsource} interactive  photometry tool of
the  Science Analysis  System (SAS),  for all  of the  above mentioned
LINER~1s in  a circular aperture  with a 3\arcsec\ radius  centered on
the source. We corrected for the background taken as an annular region
with  5\arcsec\ and  10\arcsec\ inner  and outer  radii, respectively.
These background-subtracted count rates  were then extrapolated to the
coincidence-loss  area  corresponding  to  6\arcsec\ using  the  point
spread  function. We converted  these count  rates to  physical fluxes
using        the        two        conversion        factors        of
$2.20\times10^{-15}$~erg~cm$^{-2}$~\AA$^{-1}$~counts$^{-1}$         and
$5.71\times10^{-15}$~erg~cm$^{-2}$~\AA$^{-1}$~counts$^{-1}$   for  the
UVM2                 and                 UVW2                 filters,
respectively\footnote{http://xmm.esac.esa.int/sas/current/watchout/\\Evergreen\_tips\_and\_tricks/uvflux.shtml}.

Radio data, whenever possible,  are coming from VLBA observations with
milliarcsecond   resolution.   This   is  important   to   remove  the
contamination of  any parsec- and/or subparsec-scale  jet, detected in
the nuclear region of many LINERs.  Otherwise, fluxes derived from VLA
subarcsecond  resolution were  used. Infrared  data are  only accepted
when   measured  with  the   Spitzer  telescope   with  $\sim1$~arcsec
resolution.  Even  at that  resolution, the near-IR  (1-3~$\mu$m) data
could  be highly  contaminated  by the  emission  from normal  stellar
populations.  The  mid-IR (10-30~$\mu$m), on the other  hand, are more
representative  of the  nuclear emission,  although emission  from hot
dust grains  ($\sim$100~K) can  potentially contribute to  the nuclear
flux.  Optical  data are only  taken from the \hst\  with subarcsecond
resolution.   We complement  the  SEDs with  more  \hst\ optical  data
points     by    looking    at     the    Hubble     Legacy    Archive
(HLA)\footnote{http://hla.stsci.edu/hlaview.html} for any observations
with  either  the WFPC2  or  the ACS  instruments  that  were not  yet
reported in the  literature. We use the flux  derived from the Virtual
Observatory DAOPhot tool in a 0.3\arcsec\ circular aperture around the
LINER~1,   originally   in   counts~s$^{-1}$,   and  convert   it   to
erg~cm$^{-2}$~s$^{-1}$~\AA$^{-1}$ by using the PHOTFLAM keyword in the
image  FITS file.  These  HLA photometry  data are  background, charge
transfer              efficiency,             and             aperture
corrected\footnote{http://hla.stsci.edu/hla\_faq.html}.        Finally,
Infrared,  optical,  and UV  data  are  de-reddened  according to  the
galactic  extinctions  shown  in  Table~\ref{alphaOX-Ruv-Rx-tab},  and
using the \citet{cardelli89apj:gal.ext} galactic extinction curve.

None of  the two  OM UV filters  encloses the 2500~\AA\  wavelength in
order   to    calculate   the    optical   to   X-ray    flux   ratio,
\alphaox\       \citep[\alphaox$=0.384~\log(L_{\rm       2~keV}/L_{\rm
    2500~\AA})$,][]{tananbaum79ApJ}.     Therefore,   we    derive   a
2500~\AA\ monochromatic  luminosity from either  the UVM2 or  the UVW2
fluxes, assuming a UV spectral index of 0.7 ($F_\nu\propto\nu^\alpha$,
where $F_\nu$  is in erg~s$^{-1}$~cm$^{-2}$).   This UV-spectral index
was calculated  for NGC~315, which is  the only source  to be observed
with both the UVM2 and UVW2  filters.  In fact, assuming a UV spectral
index between 0.4 and 1.0 would only introduce a maximum of 6\%\ error
on \alphaox.  Table~\ref{ngcall-sed-tab} gives the data points used to
construct the SED  of the six sources, with  simultaneous UV and X-ray
fluxes, shown  in Fig.~\ref{our-sed-liner1s}.  Finally,  we calculated
the bolometric  luminosities of these  six LINER~1s assuming  that two
consecutive   flux   points   define   a   power-law   of   the   form
$L_{\nu}\propto\nu^{\alpha}$.  These bolometric luminosities  might be
slightly biased by the fact that we know little about LINER sources in
the infra-red domain (Fig.~\ref{our-sed-liner1s}).  However, we expect
faint nuclear emission in this energy range in LINER sources since the
torus is probably absent (\citealt{ho08aa:review}, Paper~1).

\subsubsection{NGC~315}

This  immense radio  galaxy has  been the  focal point  of  many radio
studies  in  the  past \citep{willis81AA:ngc315,  venturi93apj:ngc315,
  cotton99apj:ngc315, canvin05mnras:ngc315, laing06mnras:ngc315}.  The
radio morphology  of NGC~315 indicates a Fanaroff-Riley  type 1 (FR~I)
nucleus  with  jet structures  at  both  arcsecond and  milliarcsecond
resolutions.   \citet{willis81AA:ngc315} present  arcsecond resolution
observations at 49 and 21~cm  of this large radio galaxy.  They derive
a   49--21~cm   constant   jet   spectral  index   of   $\alpha=-0.6$.
\citet{venturi93apj:ngc315} studied NGC~315 at 18, 6, and 3.6~cm using
VLBI networks  and found the same  spectral index as  above between 18
and   6~cm,  but   it  steepens   at  lower   wavelengths   (see  also
\citealt{laing06mnras:ngc315}).   \citet{cotton99apj:ngc315} explained
the sidedness  asymmetry in  the parse-scale jet  in terms  of Doppler
beaming   from   an    accelerating   relativistic   jet   (see   also
\citealt{canvin05mnras:ngc315}).         In        the       infrared,
\citet{gu07ApJ:ngc315} studied the {\sl  Spitzer} images of NGC~315 at
3.6,  4.5, 5.8,  and 8.0~$\mu$m.   The {\sl  Spitzer} telescope  has a
2\arcsec\ resolution  in these  bands, and therefore  contamination of
the nuclear fluxes by dust  emission or even red stars are inevitable.
The authors  corrected for  the background, which  basically represent
the  stellar component,  relying on  the 3.6~$\mu$m  image  (which was
shown as a good tracer  of the stellar mass distribution in elliptical
galaxies,   \citealt{pahre04ApJS:midIR})    between   10\arcsec\   and
30\arcsec.   The  real  challenge  is  therefore to  correct  for  the
heated-dust emission,  coming most certainly from the  dusty ring seen
in the optical  band (see below).  \citet{gu07ApJ:ngc315} demonstrated
that the AGN is definitely responsible for some of the emission in the
mid-IR band, since its shape is very close to the mid-IR {\sl Spitzer}
PSF  shape of  a point-like  source.  However,  with the  current {\sl
  Spitzer} data set  it is difficult to disentangle  the dust from the
AGN emission.  These  IR fluxes should then be  treated with cautious.
At optical wavelength, \hst\ images  of NGC~315 show a nuclear compact
source surrounded by  a nuclear dust disk and  chaotic dusty filaments
\citep{gonzalezdelgado08}.  We  calculated two UV  fluxes, coming from
the  UVW2 and  UVM2 filters,  which  are simultaneous  with the  X-ray
\xmm\ data.   The different fluxes  used to construct the  NGC~315 SED
(Fig.~\ref{our-sed-liner1s})          are          reported         in
Table~\ref{ngcall-sed-tab}.   The blue  line at  X-rays  represent the
\chandra, obs.  ID: 4156, fluxes and is only shown for comparison.

We  calculate  a  bolometric  luminosity using  the  \xmm\  absorption
corrected     luminosities     and     the    data     reported     in
Table~\ref{ngcall-sed-tab},  only excluding the  IR fluxes.   We found
$L_{\rm  bol}=4.34\times10^{42}$~erg~s$^{-1}$.  This corresponds  to a
X-ray  bolometric   correction  $\kappa_{2-10~keV}=L_{\rm  bol}/L_{\rm
  2-10~keV}=18.3$.  We calculate  a monochromatic 2500~\AA\ luminosity
of about  $3.2\times10^{26}$~erg~s$^{-1}$~Hz$^{-1}$.  We derive, using
this    luminosity,     an    optical    to     X-ray    flux    ratio
\alphaox$=-1.15\pm0.05$.

\subsubsection{NGC~3226}

\begin{table*}[!t]
\caption{Multiwavelength nucleus data for our sample of LINER~1s.}
\label{ngcall-sed-tab}
\newcommand\T{\rule{0pt}{2.6ex}}
\newcommand\B{\rule[-1.2ex]{0pt}{0pt}}
\begin{center}{
\resizebox{\textwidth}{!}{
\begin{tabular}{l c c c c c}
\hline
\hline
$\nu$ \T \B & $\log(\nu L_{\nu})$ & Aperture & Satellite/Filter & Date & Reference \\
(Hz)  \T \B &  (ergs~s$^{-1}$)   & (arcsec) &                  &      &           \\
\hline
\multicolumn{6}{c}{NGC~315} \T \B\\
\hline
1.41E+15 & 41.5302 & 3                   & \xmm/UVW2           & 2005 July 02     & This work \\
1.30E+15 & 41.5514 & 3                   & \xmm/UVM2           & 2005 July 02     & This work \\
5.51E+14 & 41.2664 & 0.2                 & \hst/F555W          & 1998 February 16 & \citet{gonzalezdelgado08} \\
5.47E+14 & 41.2052 & 0.2                 & \hst/F547M          & 1997 June 13     & \citet{gonzalezdelgado08} \\
3.75E+14 & 41.5289 & 0.2                 & \hst/F814W          & 1998 February 16 & \citet{gonzalezdelgado08} \\
8.44E+13 & 42.0245 & 2                   & {\sl Spitzer}/IRAC  & 2004 July 19     & \citet{gu07ApJ:ngc315}    \\
6.67E+13 & 42.1008 & 2                   & {\sl Spitzer}/IRAC  & 2004 July 19     & \citet{gu07ApJ:ngc315}    \\
5.23E+13 & 42.2015 & 2                   & {\sl Spitzer}/IRAC  & 2004 July 19     & \citet{gu07ApJ:ngc315}    \\
3.81E+13 & 42.3560 & 2                   & {\sl Spitzer}/IRAC  & 2004 July 19     & \citet{gu07ApJ:ngc315}    \\
1.50E+10 & 40.8043 & 0.005               & VLBA                & 1995 April 7     & \citet{kovalev05AJ:radioagn} \\
8.42E+09 & 40.3053 & 0.002$\times$0.005  & VLBI                & 1990 November    & \citet{venturi93apj:ngc315}  \\
5.00E+09 & 40.0919 & 0.003$\times$0.008  & VLBI                & 1989 April       & \citet{venturi93apj:ngc315}  \\
1.66E+09 & 39.5398 & 0.008$\times$0.002  & VLBI                & 1990 September   & \citet{venturi93apj:ngc315}  \\
\hline
\multicolumn{6}{c}{NGC~3226} \T \B\\
\hline
1.30E+15 & 40.65 & 3.0   & \xmm/UVM2        &  2002 March 29    &  This work \\
5.47E+14 & 41.11 & 0.2   & \hst/F547M       &  1997 Mars 18     &  \citet{gonzalezdelgado08}      \\   
9.57E+10 & 38.73 & 7.0   & NMA/NRO          &  2002 November 28 &  \citet{doi05MNRAS:radllagn}    \\  
8.40E+09 & 37.74 & 0.004 & VLBA             &  2003 October 2   &  \citet{anderson05ApJ:radliner} \\
4.90E+09 & 37.20 & 0.005 & VLBA             &  1997 June 16     &  \citet{falcke00ApJ:VLBAliners}         \\
\hline
\multicolumn{6}{c}{NGC~3718}\T \B \\
\hline
1.41E+15 & $<$41.01  & 3.0                  & OM/UVW2   & 2004 November 4   & This work \\
6.58E+14 & 40.52     & 0.3                  & \hst/F450W & 2001 October 03   & This work \\
5.00E+14 & 41.03     & 0.3                  & \hst/F606W & 2001 October 03   & This work \\
4.55E+14 & 41.24     & 0.3                  & \hst/F658N & 2004 November 16  & This work \\
4.55E+14 & 41.46     & 0.3                  & \hst/F658N & 1997 July 17      & This work \\
3.72E+14 & 40.84     & 0.3                  & \hst/F814W & 2004 November 16  & This work \\       
2.31E+11 & 39.05     & 1$\times$1           &  PdBI     & 2000 December 20  & \citet{krips07AA:radlin} \\
1.15E+11 & 38.64     & 2$\times$2           &  PdBI     & 2000 December 20  & \citet{krips07AA:radlin} \\
1.50E+10 & 37.75     & 0.15                 &  VLA      & 1999 September 05 & \citet{nagar02aap}        \\
8.40E+09 & 37.28     & 0.3                  &  VLA      & 1999 September 05 & \citet{nagar01:radobs}    \\
5.00E+09 & 37.08     & 0.5                  &  VLA      & 1999 September 05 & \citet{nagar01:radobs}    \\
4.99E+09 & 37.09     & 0.006$\times$0.005   &  EVN      & 2003 June         & \citet{krips07AA:radlin} \\
4.90E+09 & 36.95     & 0.0022$\times$0.001.7&  VLBA     & 1999 April 01     & \citet{nagar02aap}        \\
1.63E+09 & 36.42     & 0.008$\times$0.007   &  EVN      & 2002 February     & \citet{krips07AA:radlin} \\
\hline
\multicolumn{6}{c}{NGC~3998}\T \B\\
\hline
1.41E+15 & 41.27 & 3.0   & OM/UVW2    & 2001 May 9    & This work \\
1.20E+15 & 41.16 & 0.3   & \hst/F250W  & 2002 July  1  & \citet{maoz05apj:linervarUV}\\
9.08E+14 & 41.16 & 0.3   & \hst/F330W  & 2002 July  1  & \citet{maoz05apj:linervarUV}\\
5.47E+14 & 41.31 & 0.2   & \hst/F547M  & 1996 March 30 & \citet{gonzalezdelgado08}   \\
3.80E+14 & 41.00 & 0.2   & \hst/F791W  & 1996 March 30 & \citet{gonzalezdelgado08}   \\
3.68E+14 & 41.40 & 0.2   & \hst/F814W  & 2004 March 27 & \citet{capetti06AA:agn}    \\
1.50E+10 & 38.32 & 0.1   & VLA        & 1981 March    & \citet{hummel84AAradagn}   \\
8.40E+09 & 38.50 & 1.0   & VLA        & 1994 April 04 & \citet{healey07ApJS:radagn} \\
5.00E+09 & 38.52 & 0.0012 & VLBA       & 2006 May 27  & \citet{helmboldt07ApJ:vlba} \\
\hline
\multicolumn{6}{c}{NGC~4143}\T \B \\
\hline
1.36E+15 & 40.13      & 0.3                  &  \hst/F218W & 1997 May 30       & This work \\
1.41E+15 & 40.87      & 3                    &  OM/UVW2   & 2003 November 22  & This work \\
5.00E+14 & 40.93      & 0.2                  &  \hst/F606W & 2001 February 12  & \citet{gonzalezdelgado08} \\
1.50E+10 & 37.35      & 0.15                 &  VLA       & 1999 September 05 & \citet{nagar01:radobs}    \\
8.40E+09 & 37.30      & 0.15                 &  VLA       & 1999 September 05 & \citet{nagar01:radobs}    \\
5.00E+09 & 37.11      & 0.15                 &  VLA       & 1999 September 05 & \citet{nagar01:radobs}    \\
5.00E+09 & 37.12      & 0.0023$\times$0.0012 &  VLBA      & 1999 April 1      & \citet{nagar02aap}        \\
5.00E+09 & 36.18      & 0.071$\times$0.041   &  MERLIN    & 2001 October 18   & \citet{filho06AA:radio}  \\
\hline
\end{tabular}}}
\end{center}
\end{table*}

\begin{figure*}[!t]
\begin{center}
\includegraphics[angle=0,totalheight=0.20\textheight,width=0.49\textwidth]{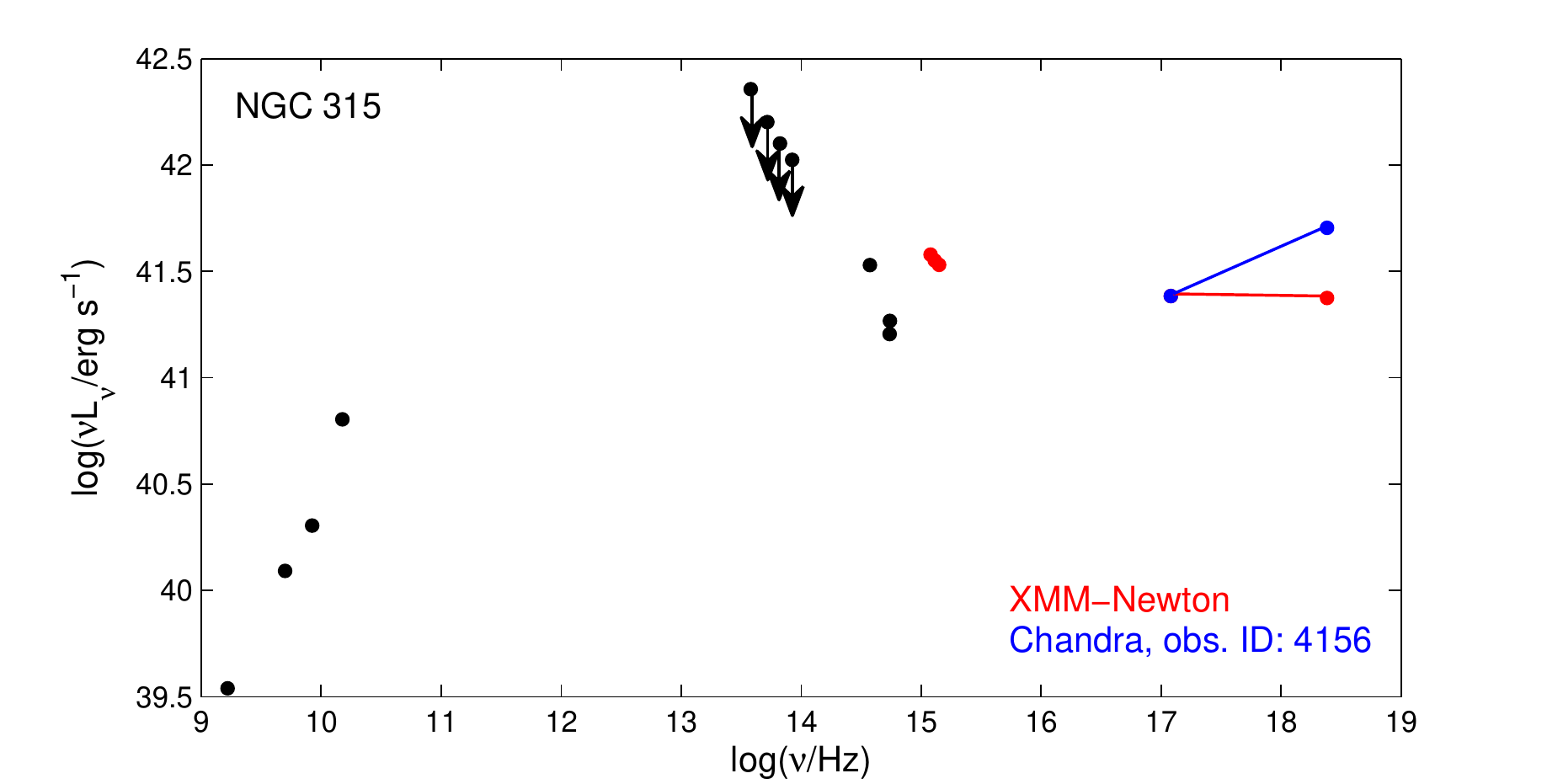}
\includegraphics[angle=0,totalheight=0.20\textheight,width=0.49\textwidth]{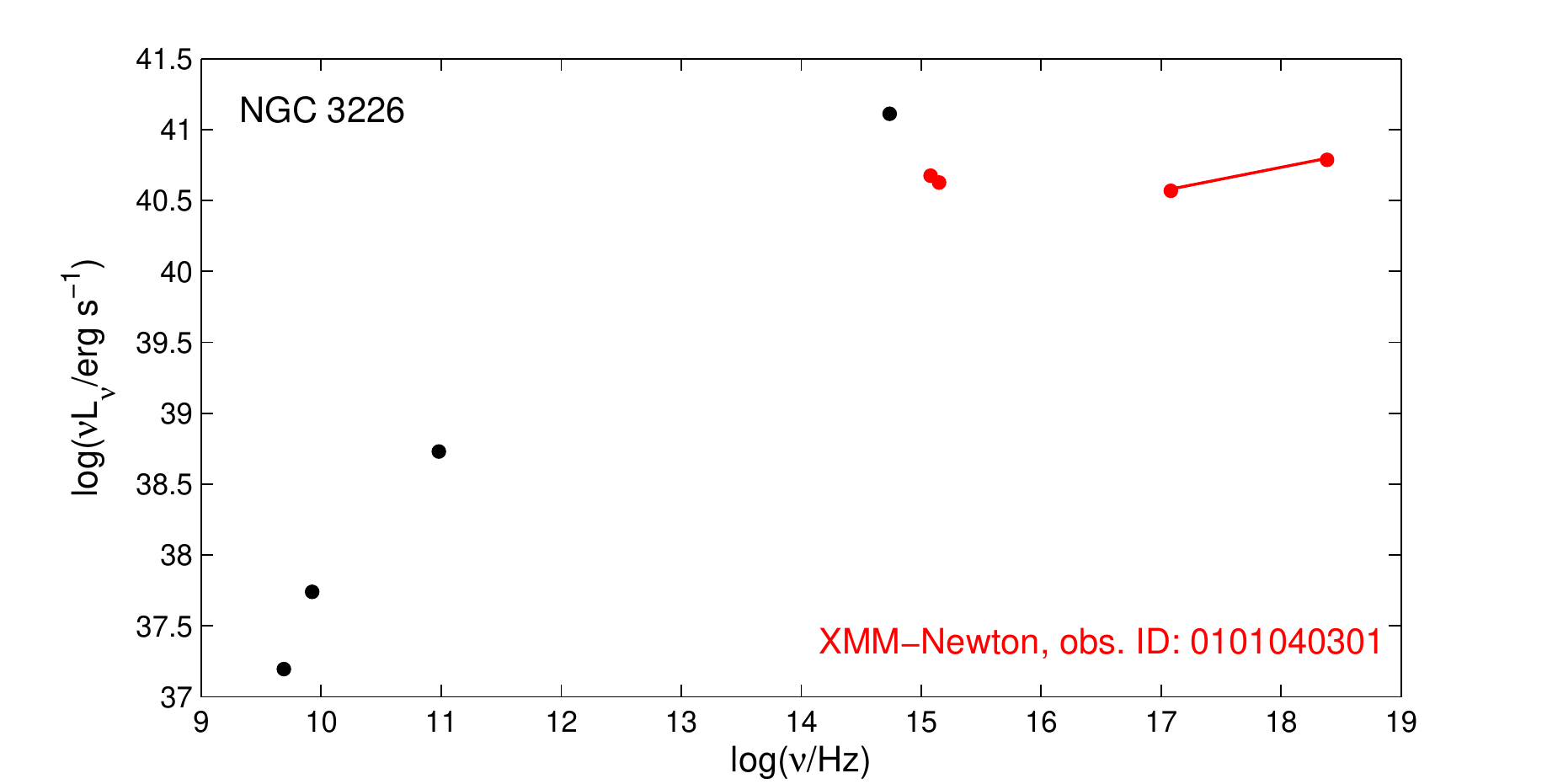}\\
\includegraphics[angle=0,totalheight=0.20\textheight,width=0.49\textwidth]{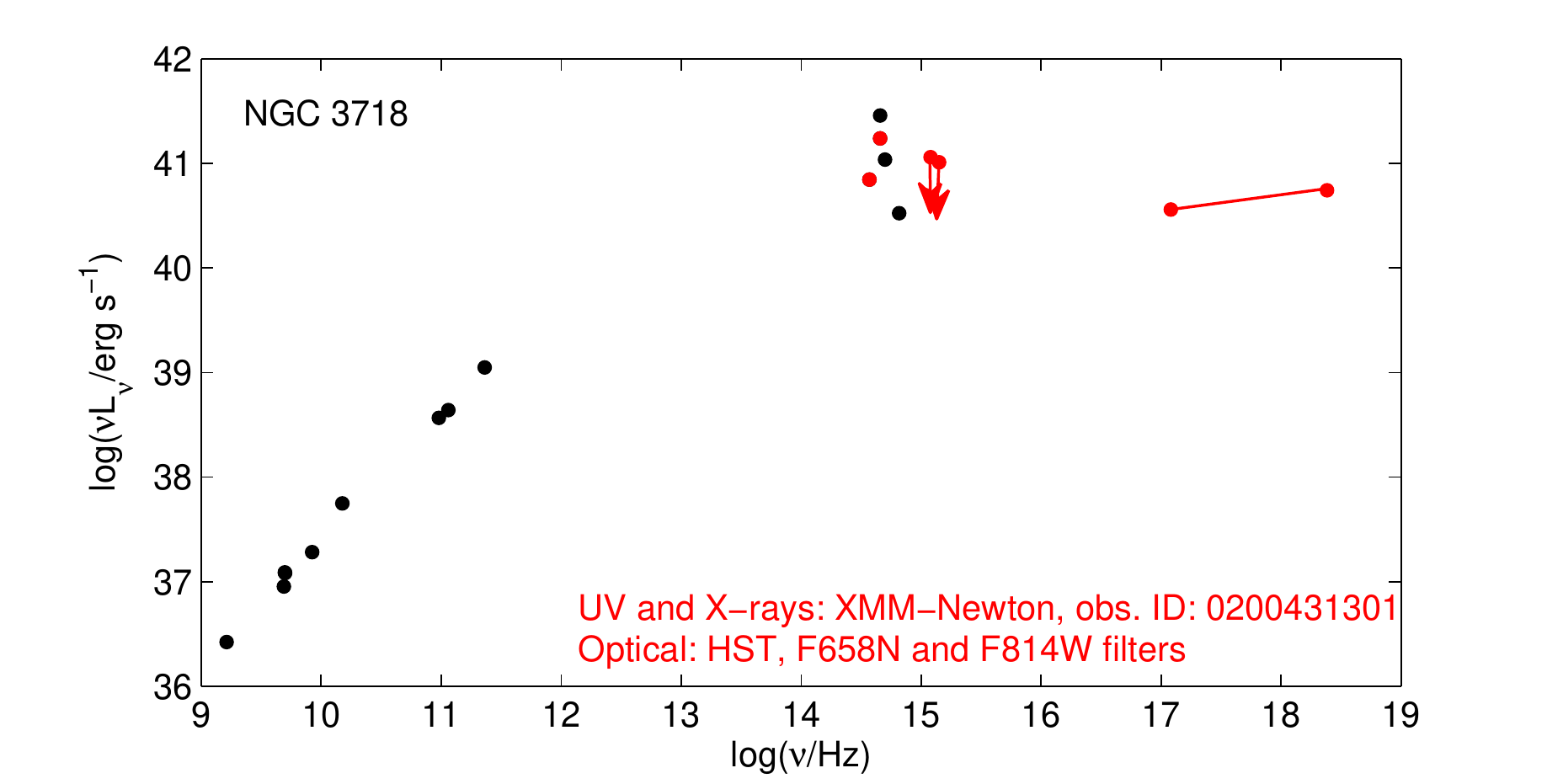}
\includegraphics[angle=0,totalheight=0.20\textheight,width=0.49\textwidth]{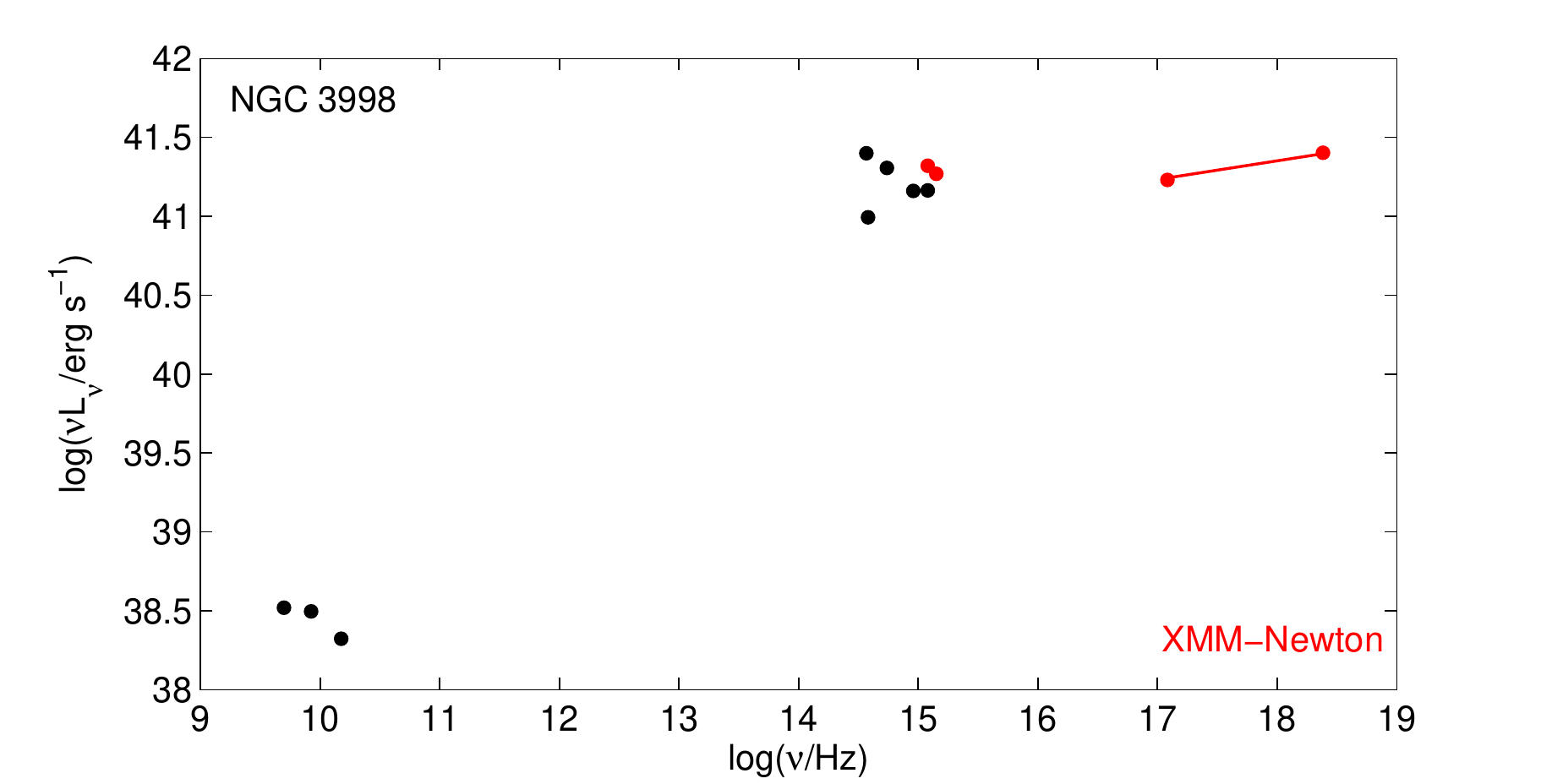}\\
\includegraphics[angle=0,totalheight=0.20\textheight,width=0.49\textwidth]{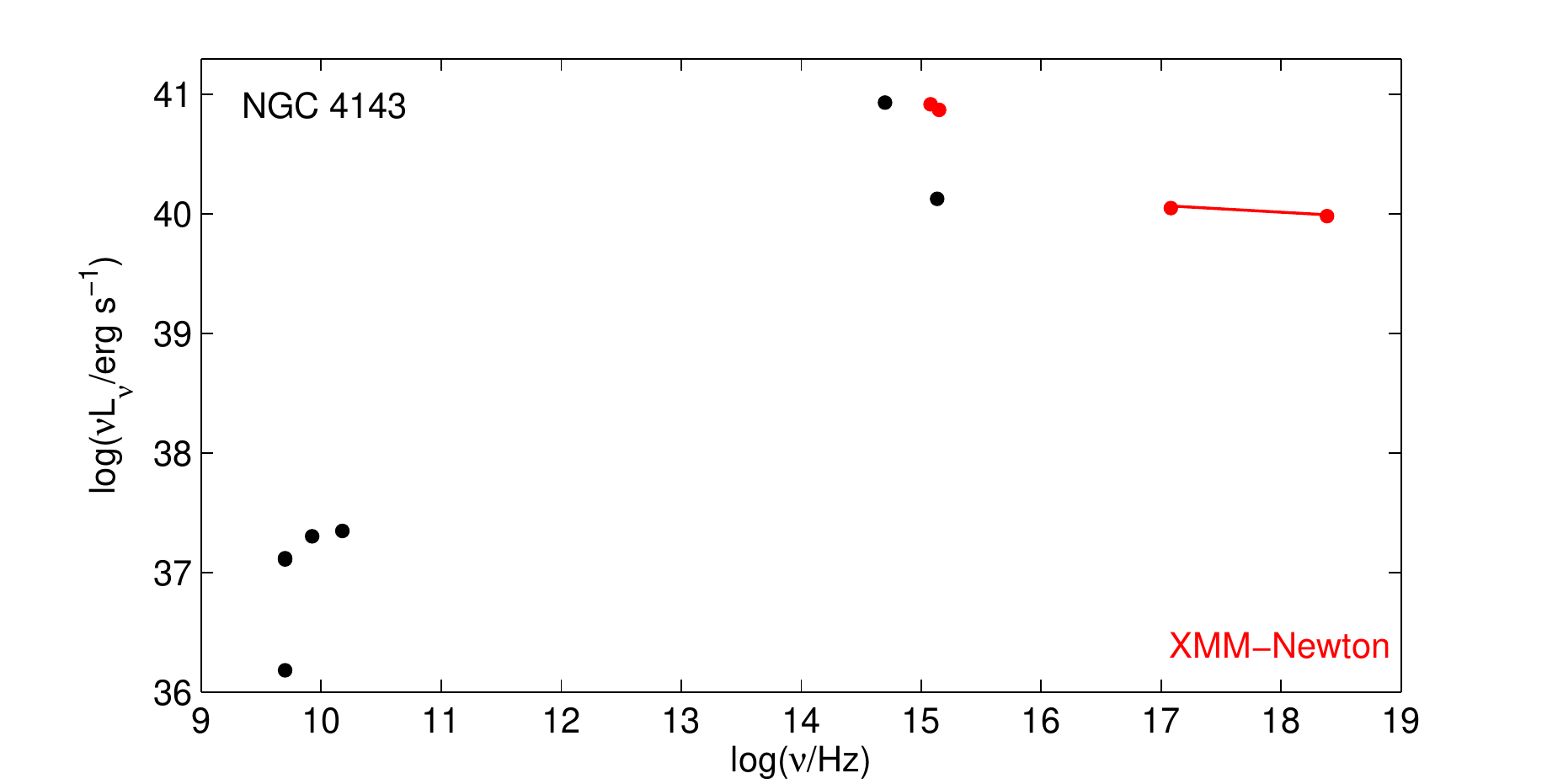}
\includegraphics[angle=0,totalheight=0.20\textheight,width=0.49\textwidth]{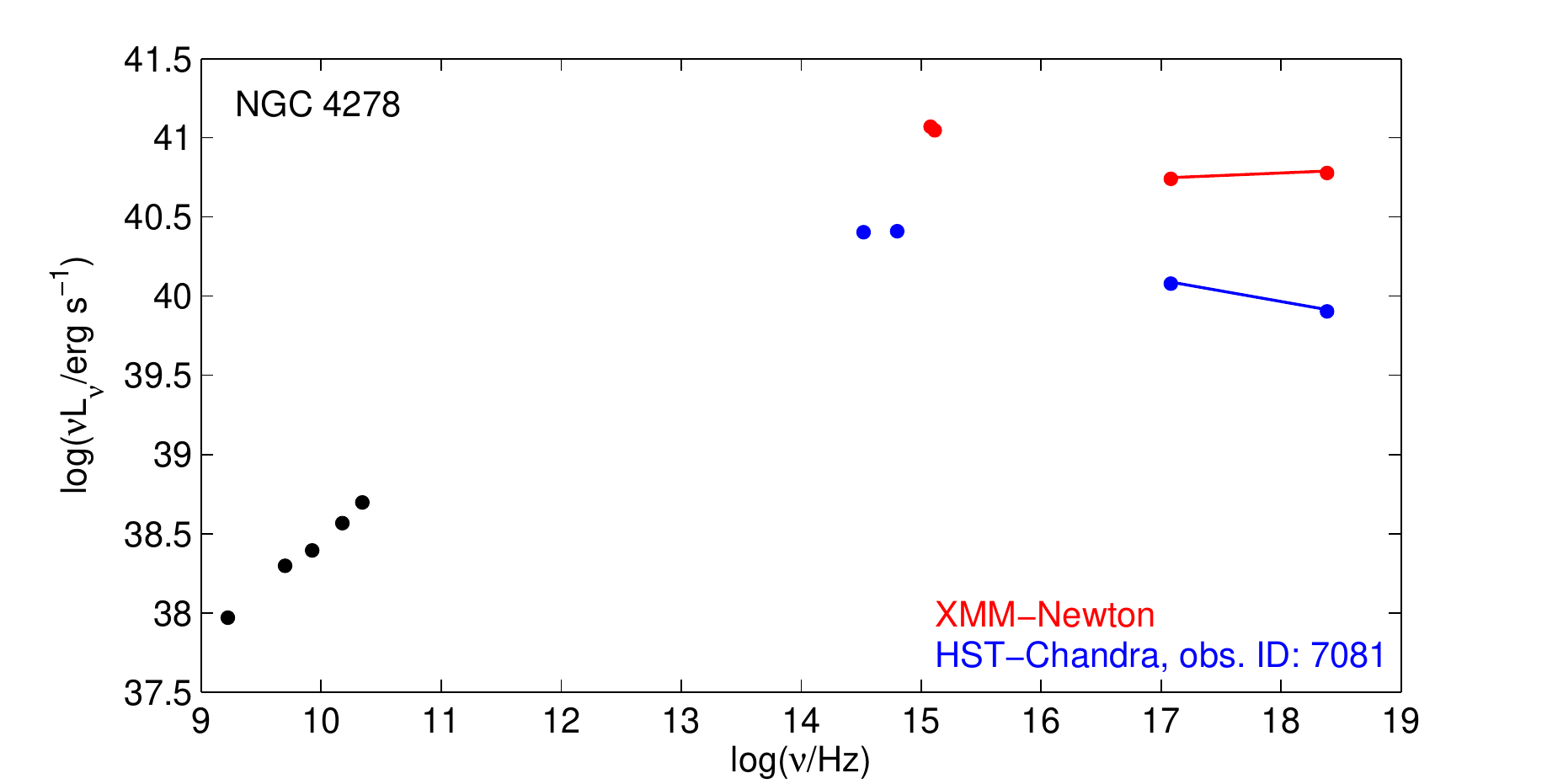}
\caption[SEDs of LINER~1s]{{\sl Top left panel.} The NGC~315 SED. The red dots represent the simultaneous \xmm\ UVM2, UVW2, and X-ray luminosities. The blue dots represent the \chandra,  obs. ID: 4156, soft and hard band luminosities. Notice the spectral and flux variability in the X-ray band (Paper~1). {\sl Top right panel.} The NGC~3226 SED. The red dots represent the simultaneous \xmm\ UV and X-ray luminosities. {\sl Middle left panel.} The NGC~3718 SED, which is one of the most complete SED of our sample with quasi-simultaneous \hst-optical and \xmm\ UV and X-ray fluxes (red dots). {\sl Middle right panel.} The NGC~3998 SED with the red dots representing the simultaneous \xmm\ UV and X-ray luminosities. {\sl Lower left panel.} The NGC~4143 SED. The red dots represent the simultaneous \xmm\ UV and X-ray fluxes. Note the high variability amplitude at UV wavelengths between the \hst\ black-dot and the OM red-dots taken six years apart. This LINER~1 shows the highest UV flux relative to the X-ray in our sample. {\sl Lower right panel.} The two NGC~4278 SEDs, representing the low state contemporary \hst-optical and \chandra-X-ray fluxes (blue dots), and the high state simultaneous \xmm\ UV and X-ray fluxes (red dots). See text for more details.}
\label{our-sed-liner1s}
\end{center}
\end{figure*}

In the  radio band, NGC~3226 is  detected with the  VLBI techniques at
milliarcsecond  resolution, most  likely  to be  a compact  point-like
source  \citep{falcke00ApJ:VLBAliners}.  \citet{nagar00apj:radioliner}
calculated a VLA  flux density of 5.4~mJy at 2~cm,  which results in a
flat  radio spectrum  between 15  and  5~GHz.  \citet{filho06AA:radio}
detected NGC~3226 with MERLIN at subarcsec scale and found a peak flux
density at  5~GHz of about  3.5~mJy, comparable to the  value reported
for  the   VLBA  data  (see   Table~\ref{ngcall-sed-tab}),  therefore,
\citet{filho06AA:radio}  concluded  that   NGC~3226  does  not  appear
variable  at 5~GHz.   In the  optical  band at  the \hst\  resolution,
NGC~3226 shows an unresolved core  surrounded from the eastern side by
a dust lane  \citep{martel04AJ:hstliner,gonzalezdelgado08}.  In the UV
band, NGC~3226 was observed with the XMM-Newton OM instrument with the
UVW2 band during only the short observation (obs.  ID: 0101040301, see
Paper~1,  the long $\sim$100~ks  observation only  used the  U filter,
which is  highly contaminated by  the host galaxy light).   The fluxes
used  to construct the  NGC~3226 SED  (Fig.~\ref{our-sed-liner1s}) are
listed in Table~\ref{ngcall-sed-tab}.

The bolometric  luminosity of NGC~3226, derived  with the luminosities
listed in Table~\ref{ngcall-sed-tab} plus the \xmm\ X-ray luminosity,
is about $5.1\times10^{41}$~erg~s$^{-1}$.   The 2-10~keV luminosity of
the \xmm\  (obs.  ID: 0101040301) represents 12\%\  of this bolometric
luminosity   and   results  in   a   2-10~keV  bolometric   correction
$\kappa_{2-10~keV}\approx8.4$.  From  the derived 2500~\AA\ luminosity
of  about   $4\times10^{25}$~erg~s$^{-1}$~Hz$^{-1}$,  we  calculate  a
$\alpha_{\rm ox}=-1.13\pm0.04$.

\subsubsection{NGC~3718}

At 18~cm, with subarcsecond  resolution of the MERLIN radio telescope,
NGC~3718  shows a  compact jet  extending to  the  northwest direction
(0.5\arcsec),    which   is   weakly    present   at    6~cm   \citep{
  krips07AA:radlin}.   The core  has  a very  high surface  brightness
temperature with  a lower limit  of $3\times10^{8}$~K. It  is detected
with  the  VLA and  VLBA  at  subarcsecond  and milliarcsecond  scales
\citep{nagar02aap}.  In  the optical  band, NGC~3718 shows  an unusual
strongly-warped      gas      disk     \citep{      pott04A&A:ngc3718,
  krips05A&A:ngc3718,   sparke09AJ:ngc3718}.    This   gas   disk   is
consistent with  the dust lane that  goes across the  nucleus and seen
clearly at the \hst\ resolution. This  dust lane makes it very hard to
detect the nucleus at optical wavelength and even more in the UV band.
In fact,  \hst\ UV images of NGC~3718  were blank with no  sign of the
nucleus  to  be  found  \citep{maoz96ApJS:uvliner,barth98ApJ:uvliner}.
This  dust  lane  is   probably  responsible  for  the  high  internal
extinction  of about  0.34~mag  \citep{ho97apjs:specparam}. Therefore,
all  of the  optical  and UV  fluxes  of NGC~3718  were corrected,  in
addition to  the Galactic extinction,  for the internal one  using the
\citet{cardelli89apj:gal.ext} curve.

We calculated optical fluxes coming from \hst\ observations, using the
fluxes reported by the  DAOPHOT Virtual-Observatory tool of the Hubble
Legacy Archive  (see above).   In the UV  band, NGC~3718  was observed
with the OM  filter UVW2 during both \xmm\  observations.  The nucleus
was not detected so we derived upper limits and corrected these fluxes
for  Galactic  and  internal  extinction.   Table~\ref{ngcall-sed-tab}
shows  the   fluxes  used  to   derive  the  NGC~3718  SED   shown  in
Fig.~\ref{our-sed-liner1s}. Luckily,  two \hst\ pointings  of NGC~3718
were  taken  almost simultaneously  with  the  \xmm\ observation  with
almost two weeks separation.  These quasi-simultaneous data points are
shown in red in Fig.~\ref{our-sed-liner1s}.

The NGC~3718 is the most  complete SED of our sample with simultaneous
data    in   the   optical    (\hst/F658N/F814W),   UV,    and   X-ray
(\xmm/0200431301)  bands.   However,  we  decided to  exclude  the  UV
upper-limit  fluxes to  calculate  the bolometric  luminosity of  this
source, hence, relying only on  the detection measurements of the radio
and quasi-simultaneous optical and X-ray fluxes (we note here that the
UV-flux upper limit is  not conflicting with the interpolation between
the  optical  and X-ray  fluxes).   we  find  a bolometric  luminosity
$L_{\rm     bol}=5.43\times10^{41}$~erg~s$^{-1}$,     resulting     in
$\kappa_{\rm  2-10~keV}=10$.    We  derive  an  upper   limit  on  the
2500~\AA\               luminosity               of              about
$9.5\times10^{25}$~erg~s$^{-1}$~Hz$^{-1}$,   which   resulted   in   a
$\alpha_{\rm ox}>-1.26$.

\subsubsection{NGC~3998}

\citet{hummel84AAradagn} reported a variable  radio core at the center
of NGC~3998  with a flat spectrum  and a very  high surface brightness
temperature     ($>10^8$~K).      At    milliarcsecond     resolution,
\citet{filho02AA:radliner}  reported a  weak northern  radio extension
suspected  to be  the innermost  kpc-scale outflow  seen  at arcsecond
resolution     \citep{hummel80AA:radagn}.     At     UV    wavelength,
\citet{maoz05apj:linervarUV} reported a strong variability between the
5 epoch observations of NGC~3998  with the \hst/F250W filter. The flux
of  the first  observation was  found the  highest and  decreasing all
along the 4  other observations.  Our \xmm/OM measurement,  which is a
$\sim1$~year before  the first \hst\ observation, is  1.3 times higher
(note that our derived UVW2 flux is consistent with the one derived in
\citealt{ptak04apj:ngc3998}  for the  same  observation).  At  optical
wavelength with the \hst\ resolution, the nucleus of NGC~3998 is clean
as a whistle,  not showing any sign of dust lanes  or any other source
of        intrinsic        absorption       \citep{gonzalezdelgado08}.
Table~\ref{ngcall-sed-tab} gives the data points used to construct the
NGC~3998 SED plotted in Fig.~\ref{our-sed-liner1s}.

Using the  different radio and  optical measurements and the  \xmm\ UV
and  X-ray  fluxes  we  calculate  a bolometric  luminosity  of  about
$1.5\times10^{42}$~erg~s$^{-1}$.   The  hard  2-10~keV  luminosity  is
17\%\  of the bolometric  luminosity which  results in  a $\kappa_{\rm
  2-10~keV}\approx6$.   From  the  derived  2500~\AA\  luminosity,  we
calculate a $\alpha_{\rm ox}=-1.05\pm0.01$.

\subsubsection{NGC~4143}

NGC~4143 has a slightly inverted  to flat radio spectrum between 20~cm
and  2~cm \citep{nagar00apj:radioliner}. At  the MERLIN  resolution of
sub-arcsecond,  NGC~4143 shows a  core emission,  lacking any  sign of
extended  jet  emission.   The  core  has a  high  surface  brightness
temperature of about  $2\times10^8$~K. In the optical band  and at the
\hst\    resolution,   \citet{chiaberge05apj:hstliner}    reports   an
unresolved nuclear  source at the  center of NGC~4143, with  some weak
nuclear  dust   lanes  \citep{gonzalezdelgado08}.   We   calculated  a
\hst/F218W UV flux for NGC~4143,  where the nucleus appears as a point
like         source.         This         flux         of        about
$3.94\times10^{-13}$~erg~cm$^{-2}$~s$^{-1}$, is  $\sim6$ times smaller
than the  flux we derive from  the OM/UVW2 observation  taken almost 6
years  later. Table~\ref{ngcall-sed-tab}  presents the  NGC~4143 flux
points used to derive the SED, shown in Fig.~\ref{our-sed-liner1s}.

Using the  different radio and  optical measurements and the  \xmm\ UV
and  X-ray  fluxes  we  calculate  a bolometric  luminosity  of  about
$3.5\times10^{41}$~erg~s$^{-1}$.   The  hard  2-10~keV  luminosity  is
3\%\  of the  bolometric luminosity  which results  in  a $\kappa_{\rm
  2-10~keV}\approx36$.   From  the  derived  2500\AA\  luminosity,  we
calculate a $\alpha_{\rm ox}=-1.37\pm0.02$, the lowest between the six
different LINER~1s with a simultaneously measured UV and X-ray flux.

\subsection{Optical to X-ray flux ratio, $\alpha_{\rm ox}$}
\label{opt-prop-sec}

We  report in Table~\ref{alphaOX-Ruv-Rx-tab}  the different  values of
$\alpha_{\rm ox}$  derived for the  six LINER~1s with  simultaneous UV
and X-ray  observations. The values  span a range between  $-1.05$ and
$-1.37$. Excluding the lower  limit calculated for NGC~3718 of $-1.27$
would  result in  a mean  of  about $-1.17\pm0.02$  with a  dispersion
$\sigma=0.01$.  These values are  comparable to the values reported in
\citet[][i.e.,  $-0.9$-$-1.2$]{maoz07MNRAS},  for a  sample  of 13  UV
variable LINERs,  and \citet{eracleous10:linersed} for  a broad sample
of  low luminosity  AGN, both  studies using  non-simultaneous  UV and
X-ray  fluxes.  However,  our  $\alpha_{\rm ox}$  values are  somewhat
lower than the mean  value reported in \citet[][ i.e., -0.9]{ho99sed}.
We  attribute this difference  to the  fact that  \citet{ho99sed} used
mainly  {\sl   ASCA}  X-ray   fluxes  derived  from   large  apertures
(300\arcsec), which would over-estimate the X-ray fluxes and hence the
$\alpha_{\rm ox}$  measurements. Moreover, the  author sample included
both LINERs and low luminosity Seyferts, with $\alpha_{\rm ox}$ values
extrapolated from the optical slope at times.

NGC~3998   is   the  only   source   our   sample   shares  with   the
\citet{maoz07MNRAS} sample.   The two $\alpha_{\rm ox}$  values are in
very  good  agreement,  separated  by  only  7\%.   NGC~4143,  with  a
$\alpha_{\rm ox}=-1.37\pm0.02$, is the  only source in our sample with
a   value   comparable   to   values   reported   for   luminous   AGN
\citep[e.g.,][]{  steffen06AJ:aox}.   This  LINER~1 has  a  comparable
2500~\AA\ luminosity to, but  the lowest 2~keV luminosity between, our
six LINER~1 sources.

\begin{figure}[!t]
\begin{center}
\includegraphics[angle=0,width=0.5\textwidth]{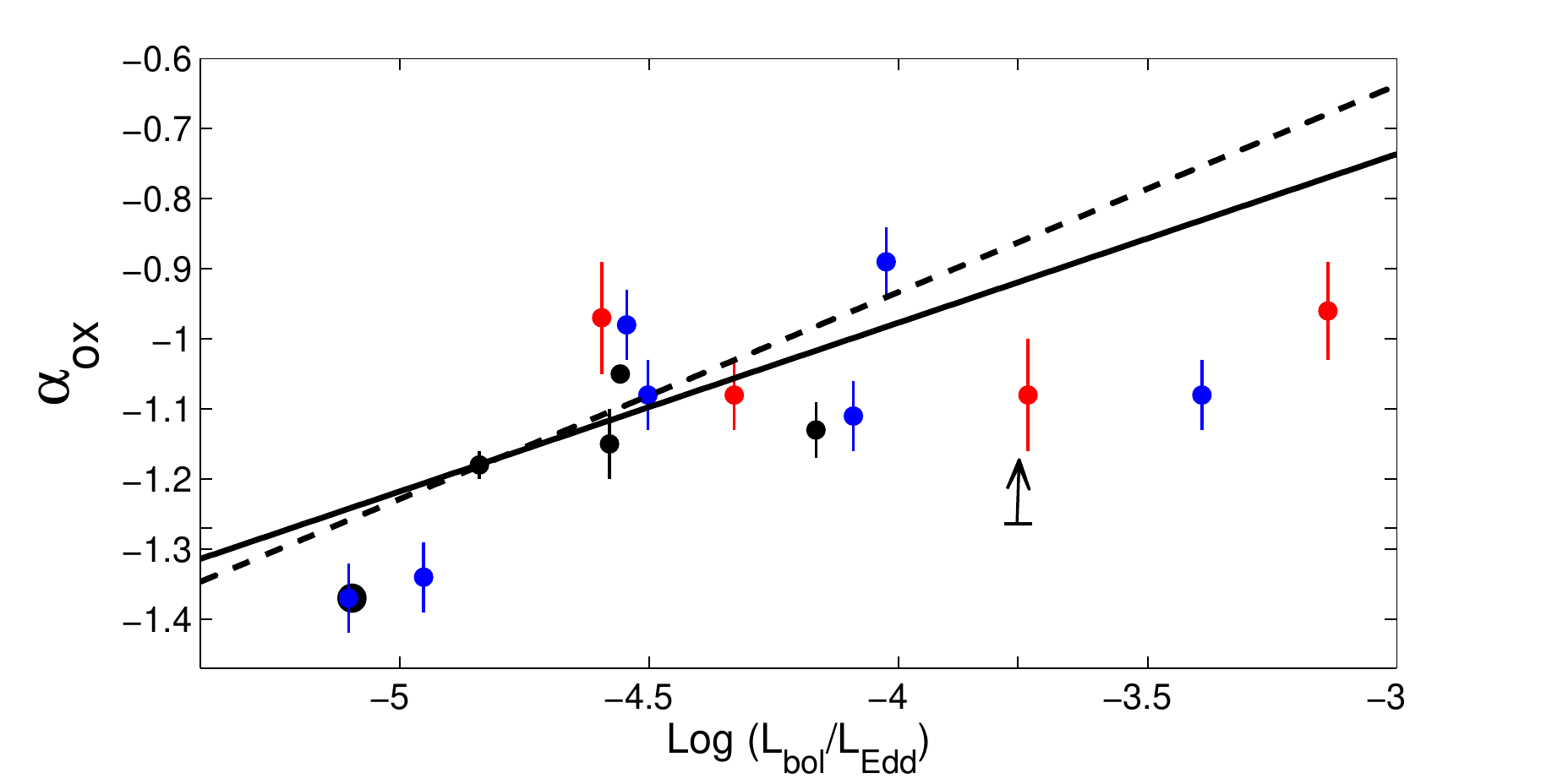}
\caption{\alphaox\ dependence on the Eddington ratio. The black-dashed line represents the best fit linear regression to our sample of LINER~1s (black dots) and the sample of \citet[][red dots]{pianmnras10}, both with \alphaox\ values derived from simultaneous UV and X-ray fluxes. The black solid-line represents the best fit linear regression to all data, including the \citet[][blue dots]{maoz07MNRAS} sample. These latter values derive from non-simultaneous observations. The upper limit represents the value derived on NGC~3718 and is not included in the regression analysis. See text for more details.}
\label{alphaOX-fig}
\end{center}
\end{figure}

Albeit  dealing with a  small sample  of six  LINER~1s, our  values of
$\alpha_{\rm  ox}$  are  not  subject  to  uncertainties  due  to  the
variability in both  UV and X-ray bands. This is  a big advantage over
the past studies conducted using non-simultaneous data. A similar work
has been carried out by  \citet{pianmnras10}, studying the UV to X-ray
flux ratio of a sample of four low luminosity AGN with simultaneous UV
and  X-ray fluxes derived  from the  {\sl Swift}  telescope. NGC~3998,
which  is  one of  the  four  sources  of \citet{pianmnras10}  sample,
exhibit a $\alpha_{\rm ox}=-0.97$, in good agreement with the value we
derive    here,    $\alpha_{\rm    ox}=-1.05$.    Recently,    \citet{
  sobolewska11MNRAS:aox}  predicted, by  simulating the  spectra  of a
sample  of AGN  based on  the evolution  pattern of  the  transient BH
binary  GRO  J1655--40, that  the  \alphaox--Eddington ratio  relation
should change signs  at $L_{\rm bol}/L_{\rm Edd}\approx10^{-2}$.  This
change of  sign could represent a  switch from a high/soft  state to a
low/hard       state       in       AGN.       \citet[][see       also
  \citealt{grupe10ApJS:agnSED}]{lusso10A&A:oxSEY}   found  that  their
sample of type 1 AGN  shows an anticorrelation (considering a negative
\alphaox)  between \alphaox\  and the  Eddington ratio.   To  test the
hypothesis    of    \citet{sobolewska11MNRAS:aox},    we    plot    in
Fig.~\ref{alphaOX-fig} our \alphaox\  values (black dots), the results
from  \citet[][red dots]{pianmnras10},  and finally  the \citet[][blue
  dots]{maoz07MNRAS}  results  against  the X-ray-corrected  Eddington
ratio  ($L_{\rm   bol}/L_{\rm  Edd}=16\times  L_{\rm  2-10~keV}/L_{\rm
  Edd}$,  see   \S\ref{sed-prop},  \citealt{ho08aa:review},  Paper~1).
This X-ray-corrected Eddington ratio  was used for consistency between
the three different  works mentioned above.  We keep  in mind that the
\citet{maoz07MNRAS}  \alphaox\  values  derive  from  non-simultaneous
observations,  and  we  excluded  the  sources with  a  2-10~keV  flux
upper-limit.    We   would   like   to   emphasize   the   fact   that
\citet{maoz07MNRAS}  and  \citet{pianmnras10}  samples of  LINERs  are
genuine  AGN, since  all show  short and/or  long term  UV variability
\citep{maoz05apj:linervarUV}.

We find a strong positive correlation (with negative \alphaox\ values)
between  \alphaox\   and  the   Eddington  ratio,  with   and  without
\citet{maoz07MNRAS}           values,          confirming          the
\citet{sobolewska11MNRAS:aox} prediction.   A Spearman-rank test gives
a 99.3\%\ probability that these  two parameters are correlated with a
correlation-coefficient    $r=0.64$    (a    probability   of    about
$\sim95$\%\ with a correlation-coefficient  $r=0.62$ if we exclude the
\citet{maoz07MNRAS} sample).  We  performed a simple linear regression
analysis weighting by the real measurement errors on $\alpha_{\rm ox}$,
and a  mean error  of 0.05 in  the case of  \citet{maoz07MNRAS} values
(derived   from   the   individual    errors   of   our   sample   and
\citet{pianmnras10} sample,  for the reason of the  lack of measurement
errors  in this  latter case),  and  found that  these two  parameters
follow the equation:

\begin{equation}
\alpha_{\rm ox}=(0.24\pm0.07)\log L_{\rm bol}/L_{\rm Edd}-(0.01\pm0.31),
\end{equation}

\noindent for  the whole sample  (solid line, Fig.~\ref{alphaOX-fig}),
and

\begin{equation}
\alpha_{\rm ox}=(0.30\pm0.10)\log L_{\rm bol}/L_{\rm Edd}+(0.25\pm0.45),
\end{equation}

\noindent considering only the  \alphaox\ derived from simultaneous UV
and X-ray fluxes (dashed line, Fig.~\ref{alphaOX-fig}).  The errors on
the slope and intercept of these relations are at the 68\%\ confidence
level  ($1\sigma$).  We  deliberately excluded  two outliers  from the
\citet{maoz07MNRAS}   sample  to   perform  this   analysis:  NGC~4552
(\eddratio$<10^{-8}$,       \alphaox$=-1.10$)       and       NGC~4594
(\eddratio$\approx10^{-7}$,  \alphaox$=-0.92$).    These  results  are
discussed in \S~\ref{sec-thin-acc-disk}.

\subsection{Radio loudness parameter, $R_{\rm X}$}
\label{rad-prop-sec}

The radio loudness parameter is usually described as the ratio between
the optical luminosity (or the UV luminosity) to the 5~GHz luminosity,
$R_{\rm  o}=L_\nu({\rm  5~GHz})/L({\rm  B})$  ($R_{\rm  UV}=L_\nu({\rm
  5~GHz})/L(2500~{\rm \AA})$).  The barrier separating radio-loud from
radio-quiet  AGN   is  defined   at  $R_{\rm  o   (UV)}=10$.   Another
alternative, better  suited to calculate the radio  loudness of LLAGN,
is  to  use  the   hard  2-10~keV  X-ray  luminosity,  $R_{\rm  X}=\nu
L_\nu({\rm   5~GHz})/L_{\rm  2-10~keV}$  \citep{terashima03apj:rloud}.
This radio loudness parameter has many advantages over its predecessor
in   the   case   of   LLAGN.    Indeed,  as   stressed   by   \citet{
  terashima03apj:rloud}, the optical  luminosity of LLAGN could easily
be contaminated by circumnuclear emission, e.g., from stars. Moreover,
one has to be careful correcting for extinction from the circumnuclear
dust present at the center of a number of low luminosity AGN. The hard
X-ray   luminosities  are   basically  insensitive   to   large  X-ray
obscuration ($N_{H}\le10^{23}$~cm$^{-2}$) and  the hard X-ray emission
is commonly believed to be the intrinsic AGN emission.  Following this
criteria,   \citet{terashima03apj:rloud}   showed   that   $\log~R_{\rm
  X}=-4.5$  would be  the barrier  separating  radio-loud ($\log~R_{\rm
  X}>-4.5$) from radio-quiet AGN.

\begin{table*}[!t]
\caption{The multiwavelength  properties of the  different LINER~1s in
  our  sample.   Columns  represent:  (1)  the galaxy  name,  (2)  the
  Galactic extinction  taken from NED,  used to correct  the infrared,
  optical and UV  data, (3) the $\alpha_{\rm ox}$  derived for the six
  LINER~1s   with  simultaneous   UV   and  X-ray   fluxes,  (4)   the
  $2500$~\AA\ luminosity  calculated from either the UVM2  or the UVW2
  \xmm-OM filters in erg~s$^{-1}$~Hz$^{-1}$,  (5) the core 5~GHz radio
  luminosity  in   erg~s$^{-1}$~Hz$^{-1}$,  (6)  the   radio  loudness
  parameter $R_{\rm X}$, (7) the Eddington luminosity (8) $\log L_{\rm
    X,   crit}/L_{\rm    Edd}=-5.356-0.17   \log(M/M_{\odot})$   which
  represents  the   critical  2-10~keV  luminosity   above  which  the
  broad-band   spectrum  of  a   given  AGN   is  thought   to  become
  jet-dominated        \citep[][see        equation       \ref{yuaneq}
  ]{yuan05apj:jetvsriaf}, (9) the  bolometric luminosity $L_{\rm bol}$
  derived from the  SED integration for each of  the six LINER~1s with
  reliable  SEDs,   and  finally  (10)  the   bolometric,  hard  X-ray,
  correction  $\kappa_{\rm  2-10~keV}$.   Two bolometric  luminosities
  were calculated in the case of NGC~4278 corresponding to the low and
  high state X-ray flux level. See text for details.}
\label{alphaOX-Ruv-Rx-tab}
\newcommand\T{\rule{0pt}{2.6ex}}
\newcommand\B{\rule[-1.2ex]{0pt}{0pt}}
\begin{center}{
\resizebox{\textwidth}{!}{
\begin{tabular}{l c c c c c c c c c}
\hline
\hline
Galaxy Name \T \B & $E(B-V)_{\rm Gal}$ & $\alpha_{\rm ox}$ & $\log(L_{2500~\rm \AA})$ & $\log(L_{\rm 5~GHz})$ & $R_{\rm X}$ & $\log(L_{\rm Edd})$ &$\log(L_{\rm X, crit}/L_{\rm Edd})$ & $\log(L_{\rm bol})$ & $\kappa_{\rm 2-10~keV}$\\
\T \B     &    & &(erg s$^{-1}$ Hz$^{-1}$) & (erg s$^{-1}$ Hz$^{-1}$) & & (erg s$^{-1}$) & & (erg s$^{-1}$) & \\
\hline
NGC~266   &  0.069  & (...)            &  (...)    &  28.08  & -2.99  & 46.54  &  -6.79  &  (...)  &  (...)  \\
NGC~315   &  0.065  & $-1.15\pm0.05$   &  26.50    &  30.39  & -1.54  & 47.16  &  -6.90  &  42.56  &  15.4   \\
NGC~2681  &  0.023  & (...)            &  (...)    &  (...)  & (...)  & 44.88  &  -6.51  &  (...)  &  (...)  \\
NGC~2787  &  0.131  & (...)            &  (...)    &  26.89  & -1.97  & 46.25  &  -6.74  &  (...)  &  (...)  \\
NGC~3226  &  0.023  & $-1.13\pm0.04$   &  25.60    &  27.50  & -3.49  & 46.16  &  -6.73  &  41.70  &  8.3    \\
NGC~3718  &  0.014  & $>$-1.27         &  $<$25.98 &  27.39  & -3.81  & 45.71  &  -6.65  &  41.73  &  9.8    \\
NGC~3998  &  0.016  & $-1.05\pm0.01$   &  26.24    &  28.82  & -2.79  & 47.17  &  -6.90  &  42.17  &  5.8    \\
NGC~4143  &  0.013  & $-1.37\pm0.02$   &  25.84    &  27.42  & -2.80  & 46.28  &  -6.75  &  41.54  &  36.3   \\
NGC~4203  &  0.012  & (...)            &  (...)    &  27.39  & -3.50  & 45.83  &  -6.67  &  (...)  &  (...)  \\
NGC~4278  &  0.029  & (...)            &  (...)    &  28.60  & -2.04  & 46.82  &  -6.84  &  41.25  &  22.24  \\
          &  0.029  & $-1.18\pm0.02$   &  25.99    &  28.60  & -2.04  & 46.82  &  -6.84  &  41.83  &  11.25  \\
NGC~4750  &  0.020  & (...)            &  (...)    &  (...)  & (...)  & 45.37  &  -6.59  &  (...)  &  (...)  \\
NGC~4772  &  0.027  & (...)            &  (...)    &  26.76  & -3.39  & 45.56  &  -6.62  &  (...)  &  (...)  \\
NGC~5005  &  0.014  & (...)            &  (...)    &  27.64  & -2.74  & 45.89  &  -6.68  &  (...)  &  (...)  \\
\hline
\end{tabular}}}
\end{center}
\end{table*}

\begin{figure*}[t]
\begin{center}
\includegraphics[angle=0,totalheight=0.20\textheight,width=0.49\textwidth]{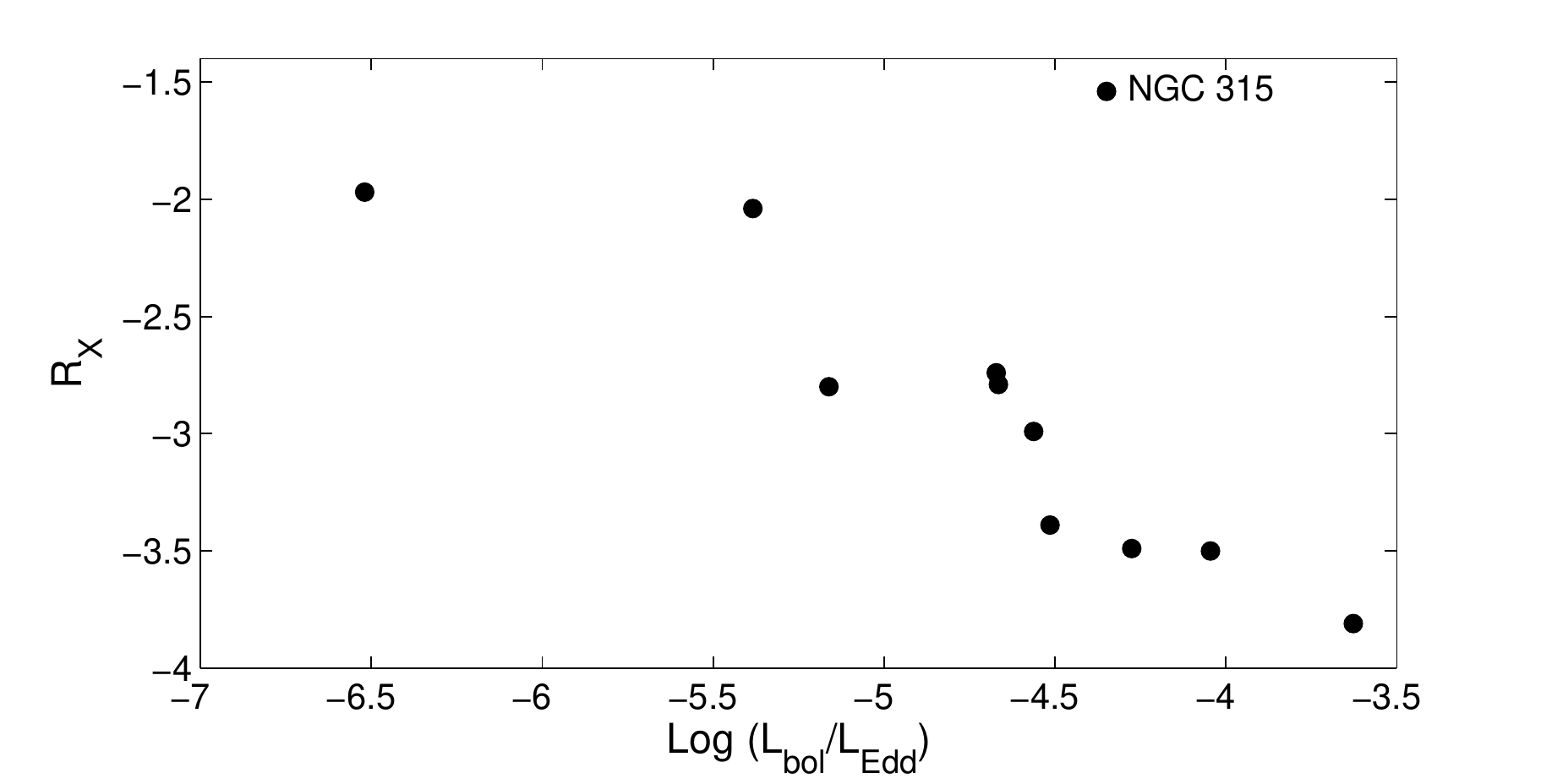}
\includegraphics[angle=0,totalheight=0.20\textheight,width=0.49\textwidth]{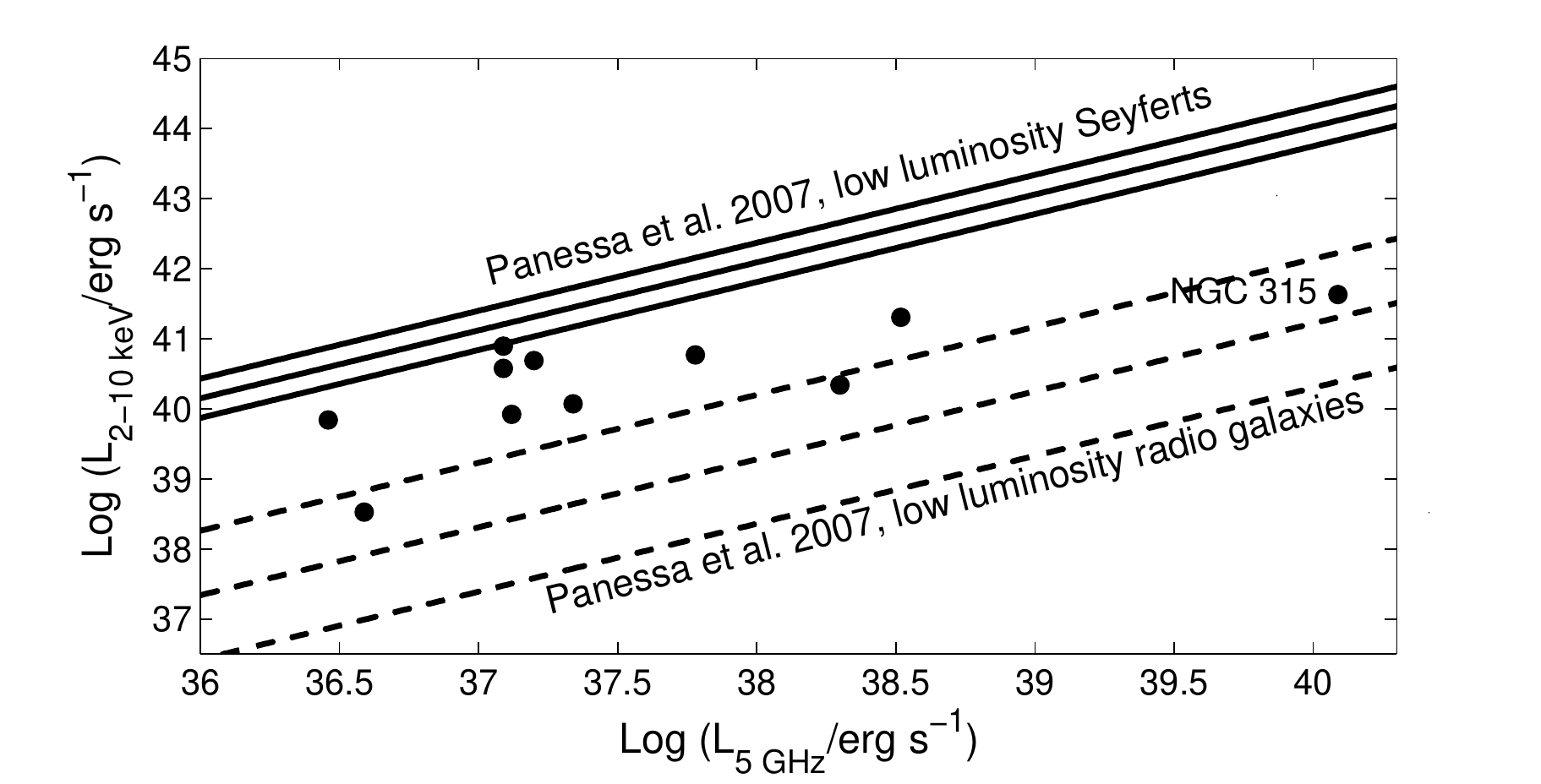}
\caption{{\sl Left panel.} The anticorrelation of $R_{\rm X}$ as a function of the Eddington ratio for the different LINER~1s in our sample. {\sl Right panel.} The positive correlation between the 5~GHz radio luminosity and the hard 2-10~keV luminosity. The solid and dashed lines represent the best fit to a straight line, and the $1\sigma$ standard deviation on the intercept, derived for a sample of low luminosity Seyfert galaxies and low luminosity radio galaxies, respectively \citep{panessa07AA:radllagn}. See text for details.}
\label{rad-loud-edd}
\end{center}
\end{figure*}

Eleven  sources in  our  sample have  5~GHz  luminosities coming  from
either  VLA  or VLBA  observations  (NGC~2681  and  NGC~4750 were  not
detected,      \citealt{4278nagar05aap}).       We      report      in
Table~\ref{alphaOX-Ruv-Rx-tab}  the different $\log~R_{\rm  X}$ values
for  the different  11  LINER~1  sources. Since  the  X-ray and  radio
observations were never simultaneous, we decided to use the arithmetic
mean   X-ray  luminosities   of  the   sources  with   multiple  X-ray
observations (Paper~1), keeping in mind that variability in both bands
may  introduce  some  scatter  on  the  derived  values.   We  find  a
$\log~R_{\rm X}>-4.5$ in all of  the cases with a geometric mean value
of $-2.7$.  According  to \citet{ terashima03apj:rloud} criterion, all
of the  11 LINER~1s  with detected 5~GHz  core could be  considered as
radio-loud sources\footnote{We note that we find a $R_{\rm UV}>40$ for
  the six sources  in our sample with derived  UV luminosities, hence,
  according to  the classical classification  criterion these LINER~1s
  would  be  considered  as  radio-loud sources.}.   However,  \citet{
  panessa07AA:radllagn}  studied a  sample of  low  luminosity Seyfert
galaxies  (type~1 and  type~2) and  a sample  of low  luminosity radio
galaxies.  They found a bimodality in the distribution of both $R_{\rm
  o}$ and $R_{\rm X}$ between the two classes and hence re-calculated,
based on the assumption that Seyfert galaxies are radio-quiet objects,
the  boundary  between  radio-loud   and  radio-quiet  sources  to  be
$\log~R_{\rm  o}>2.4$  and   $\log~R_{\rm  X}>-2.8$.   Based  on  this
assumption, 6/11  LINER~1s would be considered  as radio-loud sources.
The remaining  five sources would be  considered as intermediate-radio
sources with $\log~R_{\rm X}$  ranging between $-3$ and $-3.5$ (except
for NGC~3718 that exhibits a mean $\log~R_{\rm X}\approx-3.8$).

\citet{ho02ApJ:radLLAGN}  found a  strong anticorrelation  between the
radio-loudness  parameter $R_{\rm o}$  and the  Eddington ratio  for a
sample  of  AGN  spreading  over  almost 10  orders  of  magnitude  in
Eddington  ratio space \citep[see  also][]{sikora07apj:radloudagn}. We
show in  the left panel  of Fig.~\ref{rad-loud-edd} the  dependence of
the  radio loudness parameter,  $R_{\rm X}$,  on the  Eddington ratio.
There is clearly a  strong anticorrelation between the two parameters.
\citet{panessa07AA:radllagn}  found   the  same  anticorrelation  when
considering the $R_{\rm  X}$ parameter for a sample  of low luminosity
Seyfert galaxies and low luminosity radio galaxies. Here, we establish
this correlation for the first  time for LINER~1s.  We find a Spearman
rank   correlation  coefficient   $r=-0.71$  and   a   probability  of
98.5\%\ that  these two parameters are correlated.   NGC~315, which is
the only FR  I radio galaxy in our sample clearly  does not follow the
same  anticorrelation.   Radio galaxies  (i.e.,  FR~I radio  galaxies,
radio-loud quasars,  and broad-line  radio galaxies) follow  their own
anticorrelation      in     Eddington-ratio$-R_{\rm      o}$     space
\citep{sikora07apj:radloudagn}, and probably their own anticorrelation
in  Eddington-ratio$-R_{\rm X}$  space.  Therefore,  excluding NGC~315
from the data would  result in a Spearman-rank correlation coefficient
$r=-0.95$ and a probability  $>99.99$\%\ that these two parameters are
correlated.

\begin{figure*}[!t]
\begin{center}
\includegraphics[angle=0,width=0.80\textwidth]{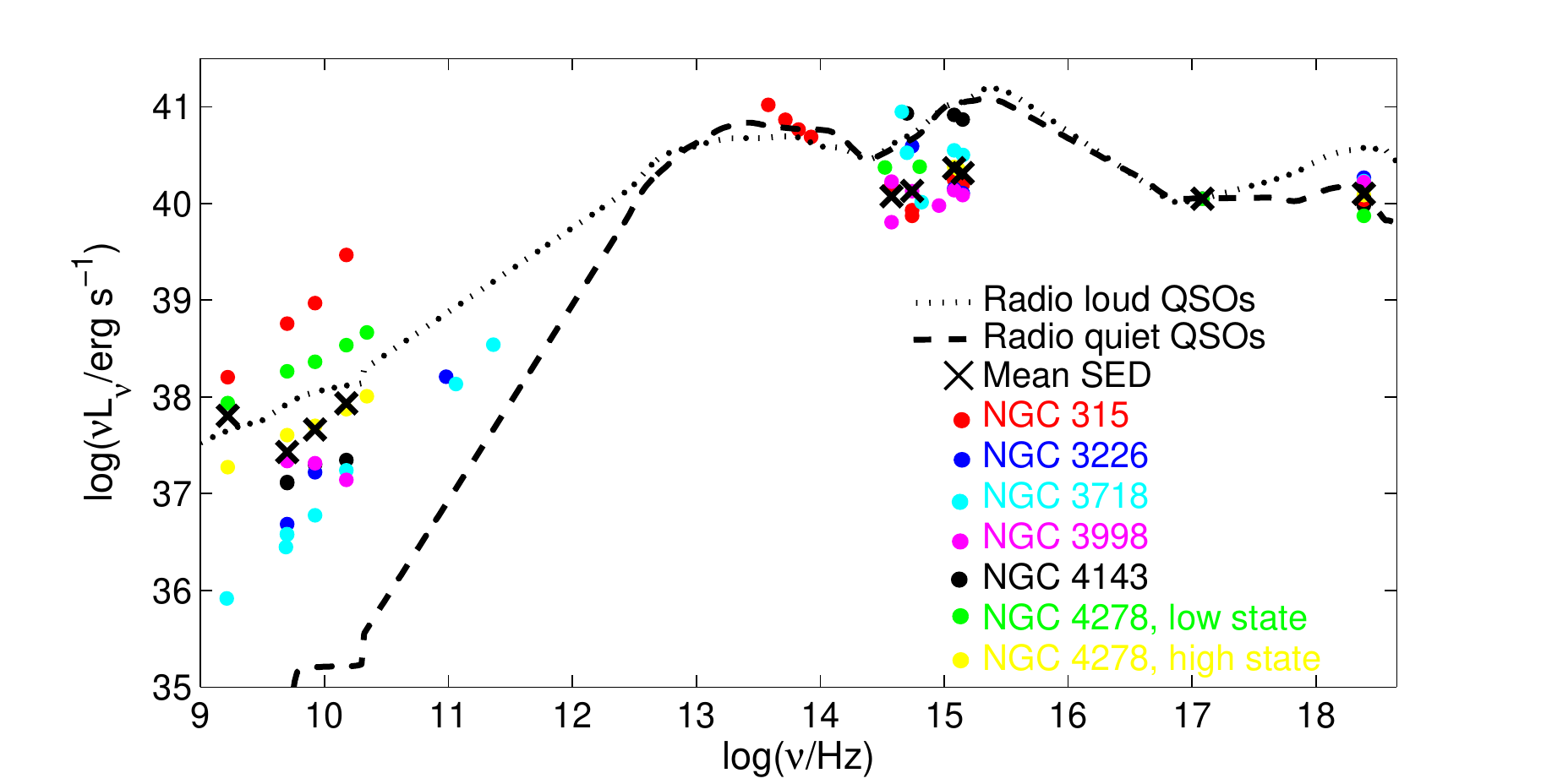}
\caption{The SEDs of the six different LINER~1s in our sample with simultaneous UV and X-ray fluxes (different colored-dots) plotted alongside their geometric mean SED (black crosses) and the mean SED of a sample of radio-quiet and radio-loud AGN \citep{elvis94apjs:quasar}. See text for more details.}
\label{mean-sed}
\end{center}
\end{figure*}

Finally,   \citet{panessa07AA:radllagn}   found   a  strong   positive
correlation  between  the  hard  2-10~keV  luminosity  and  the  radio
luminosity at  6~cm for  a sample of  low luminosity  Seyfert galaxies
\citep[see  also][]{degasperin11MNRAS:llagn}  and   a  sample  of  low
luminosity   radio  galaxies.   We   show  in   the  right   panel  of
Fig.~\ref{rad-loud-edd}  the   dependence  of  the   mean  hard  X-ray
luminosity of  our sample  of LINER~1s on  the 6~cm  radio luminosity.
According  to  the  Spearman-rank   test,  these  two  parameters  are
positively correlated with  $r=0.69$ and $p=99$\%. The best  fits to a
straight  line  derived by  \citet{panessa07AA:radllagn}  on both  low
luminosity Seyfert  and radio galaxies  are shown as solid  and dashed
lines,  respectively, with  the  $1\sigma$ standard  deviation on  the
intercepts.  Our  LINER~1 sample  clearly populate the  region between
the \citet{panessa07AA:radllagn}  two best fit  lines, indicating that
LINER~1s,  and  all  AGN-powered  LINERs by  extrapolation,  are  more
radio-loud, for a given  X-ray luminosity, than low luminosity Seyfert
galaxies and less radio-loud  than low luminosity radio galaxies.  All
these results are discussed in \S~\ref{sec-rad-jet}.

\section{Properties of LINER~1 SEDs}
\label{sed-prop}

In order to get a better  insight on our LINER~1 SEDs, compare them to
each   other   and   to    more   luminous   AGN,   we   overplot   in
Fig.~\ref{mean-sed}  the  SED of  the  six  different LINER~1s,  after
normalizing them  to the  0.5-2~keV luminosity of  NGC~4143 (different
colored-dots represent different SEDs, see figure legend).  The choice
of NGC~4143  is somewhat  arbitrary and was  preferred for  having the
lowest 0.5-2~keV luminosity (our conclusions would not, in any way, be
affected  by the  choice of  the normalizing  source).  Two  SEDs were
reported in  the NGC~4278  case, one corresponding  to the  high state
simultaneous  X-ray  and UV  \xmm\  fluxes  and  one representing  the
contemporary \hst-optical  and \chandra-X-ray (obs.   ID: 7081) fluxes
(see \citealt{younes10aa:ngc4278}).  We  decided to normalize all SEDs
preferentially at  the X-ray luminosity,  since our main focus  is the
validity of a ``big blue bump'' at UV wavelength relative to the X-ray
emission, plus to  check whether they are to  be considered radio-loud
or radio-quiet  sources relative to X-rays.   For comparison purposes,
we  plot the  \citet{elvis94apjs:quasar} average  SED of  a  sample of
radio-loud   and  radio-quiet   AGN   (dotted-line  and   dashed-line,
respectively; we  note that \citet{shang11ApJS:sedagn}  found that the
overall shape  of the average SED  of their sample  of radio-quiet and
radio-loud AGN, compiled  using high-quality multiwavelength data from
space-based  and  ground-based  telescopes,  is very  similar  to  the
\citet{elvis94apjs:quasar}  SEDs  compiled  almost two  decades  ago).
Finally, we add the  geometric\footnote{The geometric mean of a vector
  $X=[X1,X2,...,Xn]$    is    defined   as    $(\prod\limits_{i=1}^{n}
  X_i)^{1/n}$.}  mean SED (black crosses, the geometric mean minimizes
the effect  of extreme outliers)  that we calculated whenever  we have
three or  more data points of  different sources at  a given frequency
$\nu_0$.   To  make use  of  the  data as  much  as  possible, a  flux
data-point, at  radio or optical  wavelength, measured at  a frequency
$\nu_1$, in the  interval between $\nu_0-0.1$~dex and $\nu_0+0.1$~dex,
was considered as calculated at  $\nu_0$.  In other words, we consider
a source  to have a flat  spectrum in a frequency  interval of 0.2~dex
(we note  that modeling the SEDs  with accretion/jet models  is out of
the scope of this work and will be treated in a forthcoming paper).

The different  properties of the SEDs  of our sample  of LINER~1s, and
the property of the mean SED, which represents to some extent LINER~1s
as a  class and all  AGN-powered LINERs by extrapolation,  compared to
the SED of  radio-quiet and radio-loud quasars could  be summarized as
follows:

\begin{enumerate}

\item At radio wavelength and for a given X-ray luminosity, all of the
  six  LINER~1s in our  sample present  a radio  emission at  least an
  order of magnitude  larger than that of radio-quiet  quasars. Two of
  the sources, NGC~315 and NGC~4278, exhibit radio emission comparable
  to  radio-loud quasars.   The radio-loudness  parameter  $R_{\rm X}$
  indicates  that all  of  the  six LINER~1s  could  be considered  as
  radio-loud    sources    having    $\log   R_{\rm    X}>-4.5$    (see
  \S~\ref{rad-prop-sec}).

\item At a  given X-ray luminosity, the geometric-mean  optical and UV
  fluxes are 5 to 10  times, respectively, weaker than the optical and
  UV emission  of both radio-quiet and radio-loud  quasars. This leads
  to      an     optical-to-X-ray     flux      ratio     $\alpha_{\rm
    ox}\approx-1.17\pm0.02$  (see \S~\ref{opt-prop-sec}),  compared to
  mean values  of about  $-1.3$ and $-1.5$  for Seyferts  and quasars,
  respectively     \citep{mushotzky89ApJ:OXsey,     brandt00ApJ:oxAGN,
    steffen06AJ:aox, lusso10A&A:oxSEY}.   Consequently, the ``big blue
  bump'', clearly seen in the SED of quasars, is much less apparent in
  the SED of our sample of LINER~1s.

\item In  the X-ray band, the  geometric mean spectrum  in $\nu L_\nu$
  space is flat and hence comparable to those of radio-quiet AGN. This
  is quite a surprise since in the radio-band these sources have radio
  luminosities, normalized  to the X-ray band, comparable  to those of
  radio-loud quasars. In fact, the  radio loud quasars have an average
  X-ray  photon index  of about  $\Gamma\approx1.5$  \citep[e.g., ][]{
    worrall90ApJ:radAGN,    yuan98AA:radAGN,    reeves97MNRAS:quaspec,
    reeves00MNRAS:sofexc,                           gambill03AA:radAGN,
    belsole06MNRAS:radAGNxray,miller11ApJ:radAGN},    resulting   from
  either X-rays  dominated by synchrotron  emission from a jet  or the
  consequence  of X-ray  beaming  effects.  The  X-ray spectral  shape
  resemblance  that our  LINER~1s (with  an average  $\Gamma=1.9$, see
  Paper~1) share  with radio-quiet quasars, and  also Seyfert galaxies
  \citep{nandra97apj:SEYfekline,porquet04aa:pgquasar},   could   point
  toward an  accretion flow  origin rather than  a jet-origin  for the
  X-ray emission.

\item  Finally,  we  would  like  to note  here  that  the  bolometric
  luminosities of  these LINER~1s, derived from  their spectral energy
  distribution,  are extremely  low compared  to luminous  AGN.  These
  bolometric   luminosities    vary   between   $2\times10^{41}$   and
  $4\times10^{42}$,  resulting in  Eddington ratios  between $10^{-4}$
  and $3\times10^{-6}$.   The hard X-ray luminosities  present 3\%\ to
  17\%\  of these  bolometric luminosities  and result  in  an average
  2-10~keV  bolometric correction,  $\kappa_{\rm 2-10~keV}$,  of about
  16.    This    value   is   equal    to   the   one    reported   in
  \citet[][$\kappa_{\rm  2-10~keV}\approx16$]{ho08aa:review},  and  is
  smaller  than  the values  reported  for  more  luminous AGN  (i.e.,
  $\sim30$),  in  agreement  with  the anticorrelation  found  between
  $\kappa_{\rm     2-10~keV}$      and     the     Eddington     ratio
  \citep[e.g.][]{vasudevan09MNRAS:agnSED}.

\end{enumerate}

\section{Discussion}
\label{discussion-multi-corr}

 We  have shown  in Paper~1  that AGN-powered  LINERs, based  on their
 X-ray  temporal  and  spectral  properties, might  have  a  different
 accretion  mode  than the  one  thought  to  exist in  luminous  AGN.
 Briefly, we  have shown  that, (1) fast  X-ray variability  (hours to
 days)  is rare  in these  sources,  (2) the  X-ray spectral  features
 defining   most  luminous   AGN   are  absent   in  LINER~1s   (e.g.,
 Fe~K$\alpha$,  soft  excess below  $\sim$2~keV),  and  (3) the  X-ray
 photon index $\Gamma$ is anticorrelated to the Eddington ratio, which
 could be interpreted  as the increase in the optical  depth of a RIAF
 with increasing Eddington ratio leading to the hardening of the X-ray
 spectrum.   In  this work,  we  complement  Paper~1  by studying  the
 multiwavelength  properties of the  same sample  of LINER~1s.  In the
 following,  we discuss  our  results in  the  framework of  accretion
 mechanisms and  radiative processes  taking place in  LINER~1s, hence
 all  AGN-powered  LINERs  by   extrapolation.   We  seek  answers  by
 comparing  LINER~1s multiwavelength properties  to the  properties of
 luminous AGN and the behavior of transient XRBs.

\subsection{The enigma of a standard accretion disk in LINERs}
\label{sec-thin-acc-disk}

The standard geometrically thin accretion disk model emits most of the
energy    in   the   UV    in   the    case   of    a   BH    with   a
$\sim10^6-10^9$~M$_{\odot}$. This is in  line with the observations of
Seyferts and quasars, which show a ``big blue bump'' at UV wavelengths
in their  SEDs.  The bump  translates into a  UV to X-ray  flux ratio,
$\alpha_{\rm ox}$, in the range  between $-1.3$ to $-1.5$.  This UV to
X-ray flux ratio, $\alpha_{\rm ox}$,  is an essential tool in order to
understand  the link between  the UV  and X-ray  emission and  to test
accretion  mechanisms  and radiative  processes  in  AGN.  The  sparse
studies of $\alpha_{\rm ox}$  in low luminosity AGN, including LINERs,
showed  that this  class of  faint AGN  show larger  $\alpha_{\rm ox}$
values and  hence a shallower  UV ``big blue  bump''.  \citet{ho99sed}
derived an average value for $\alpha_{\rm ox}$ of about $-0.9$ for his
sample of seven low luminosity  AGN, somewhat larger than the value we
report  here  ($-1.17$).   The  UV  and X-ray  flux  ratios  were  not
simultaneous, plus,  their X-ray fluxes  were coming from  low spatial
resolution   observations  with  apertures   as  big   as  300\arcsec.
\citet[][see   also   \citealt{eracleous10:linersed}]{   maoz07MNRAS},
calculated the  $\alpha_{\rm ox}$ of a  sample of 13  LINERs (being of
type~1 or type~2) showing UV  variability on time-scales of years.  He
found  an average  value of  almost $-1.13$.   This value  is somewhat
consistent with  \citet{ho99sed} and in  very good agreement  with our
results that show a mean value of about $-1.17\pm0.02$, which is based
on simultaneous UV and X-ray observations.

\begin{figure}[!t]
\begin{center}
\includegraphics[angle=0,width=0.5\textwidth]{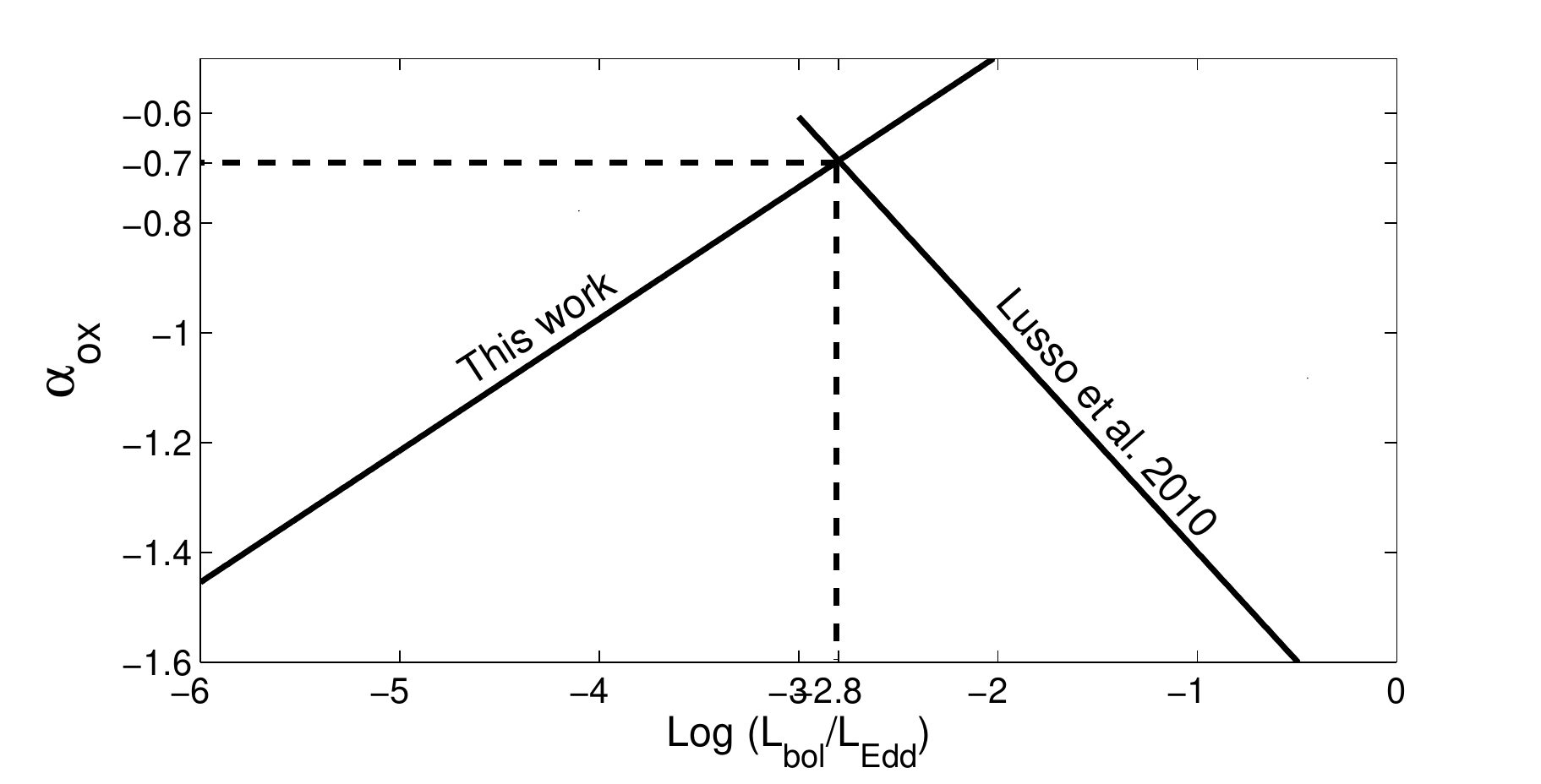}
\caption{Positive and negative correlations between \alphaox\ and the Eddington ratio for our sample of AGN-powered LINERs (\S~\ref{opt-prop-sec}) and a sample of type~1 AGN \citep{lusso10A&A:oxSEY}, respectively. The crosspoint at $L_{\rm bol}/L_{\rm Edd}\approx10^{-2.8}$ might suggest a transition from a high/soft state AGN where a thin accretion disk exists to a low/hard state LLAGN where a RIAF forms.}
\label{aOX-state-switch}
\end{center}
\end{figure}

Many studies  in the  past showed that  $\alpha_{\rm ox}$  is strongly
anticorrelated to the  monochromatic 2500~\AA\ luminosity in classical
AGN   \citep{zamorani81ApJ:OXquas,   avni82ApJ:aox,  vignali03AJ:Aox,
  strateva05AJ:OXsey,         steffen06AJ:aox,        kelly07ApJ:quas,
  lusso10A&A:oxSEY,   kelly10ApJ:quasar,   grupe10ApJS:agnSED}.   This
means that the  prominent UV ``big blue bump''  seen in quasars, which
result  in an  average  $\alpha_{\rm ox}\approx-1.4$  \citep[e.g.,][]{
  elvis94apjs:quasar},  tend to  vanish at  very low  UV luminosities.
\citet[][see also \citealt{pianmnras10}]{maoz07MNRAS}, by showing that
the  $\alpha_{\rm   ox}$  values  of  his  sample   follows  the  same
$\alpha_{\rm ox}-L_{\rm  2500\AA}$ anticorrelation shared  by Seyferts
and  quasars \citep{steffen06AJ:aox},  suggested that  a geometrically
thin  accretion   disk  could   survive  very  low   accretion  rates.
Additionally,  \citet{maoz07MNRAS} showed  that the  $\alpha_{\rm ox}$
value of  his sample are similar  to values reported for  AGN of X-ray
luminosities      ranging      from     $10^{41}$~erg~s$^{-1}$      to
$10^{43}$~erg~s$^{-1}$      and     having      low      BH     masses
($\sim10^{6}$~M$_{\odot}$).   These  low   BH-mass  AGN  have  a  mean
$\alpha_{\rm ox}$ of about $-1.2$ \citep{greene07apj:imagn}.

The  BH mass appears  to be  an essential  parameter in  the accretion
physics around BHs. Indeed, the hard 2-10~keV luminosity is positively
correlated to the BH mass \citep{ho09apj:riaf}, and hence $\alpha_{\rm
  ox}$ should be  BH mass dependent. \citet{kelly08ApJS:quasar} showed
that, for a large sample of 318 radio quiet quasars, $\alpha_{\rm ox}$
is strongly  anticorrelated with  the BH mass  (note here that  we are
assuming negative  $\alpha_{\rm ox}$ values, not the  other way around
as  treated  by  the authors).   They  were  able  to explain  such  a
dependence with  a UV to X-ray  spectrum inferred from  a simple model
describing the  standard geometrically thin accretion  disk. This last
anticorrelation explains the  high $\alpha_{\rm ox}$ values calculated
for the low BH-mass  AGN of \citet{greene07apj:imagn} in the framework
of a geometrically thin accretion disk.  However, the BH masses of the
samples of \citet{maoz07MNRAS}, \citet{pianmnras10}, and this work lie
in an interval  between $10^{7}$~M$_{\odot}$ and $10^{9}$~M$_{\odot}$.
These masses should result in  a $\alpha_{\rm ox}$ between $-1.24$ and
$-1.58$ \citep{kelly08ApJS:quasar}.  Fifteen  out of nineteen in these
three  samples  have $\alpha_{\rm  ox}$  greater  than $-1.2$,  hence,
inconsistent with the above interval.

Return  to  the  $\alpha_{\rm  ox}-L_{\rm  2500\AA}$  anticorrelation,
\citet{sobolewska11MNRAS:aox} showed  that, by simulating  the spectra
of a sample of AGN based  on the evolution pattern of the transient BH
binary  GRO  J1655--40,  LINERs   roughly  follow  the  dependence  of
\alphaox\ on the optical  luminosity of AGN. Nonetheless, the authors
showed  that LINERs  should  be in  a  low/hard state  to exhibit  the
observed   low   optical   luminosities,   $\log   (\nu   L_{\nu})_{\rm
  o}\approx39.5-41.3$,  again, considering  their  relatively high  BH
masses, $10^7-10^9$~M$_{\odot}$.

\citet{sobolewska11MNRAS:aox} predicted  a change  in the sign  of the
\alphaox--Eddington   ratio   relation   below   $L_{\rm   bol}/L_{\rm
  Edd}\approx10^{-2}$,  where these  two parameters  become positively
correlated  (considering   negative  \alphaox).   This   change  would
correspond  to  a switch  from  a high/soft  to  a  low/hard state  in
transient   XRBs.     We   confirm   in    \S~\ref{opt-prop-sec}   the
\citet{sobolewska11MNRAS:aox}    prediction    and    we    show    in
Fig.~\ref{alphaOX-fig}   the    positive   \alphaox--Eddington   ratio
correlation  that we  establish for  a sample  of  AGN-powered LINERs.
This correlation  is in contrast with  the one found  for luminous AGN
\citep{lusso10A&A:oxSEY,    grupe10ApJS:agnSED}.    Overplotting   our
positive    correlation   with    the    anticorrelation   found    by
\citet{lusso10A&A:oxSEY}   for   a   sample   of  type~1   AGN   (Fig.
\ref{aOX-state-switch}),  we find  that  the switch  from a  high/soft
state to a low/hard state in AGN might occur at the transitional point
corresponding  to  $L_{\rm  bol}/L_{\rm  Edd}\approx10^{-2.8}$  and  a
\alphaox\ of about $-0.7$.   This Eddington ratio limit that separates
two different  radiative behaviors between luminous AGN  and LLAGN has
been reported  in the  past using the  dependence of the  X-ray photon
index,  $\Gamma$,   on  the   Eddington  ratio.   Paper~1   (see  also
\citealt{gu09mnras:gamVSeddllagn})  predicted  that  below  a  similar
Eddington  ratio  value  ($L_{\rm bol}/L_{\rm  Edd}\approx10^{-2.6}$),
LINER~1s might have a RIAF  accretion flow instead of a thin accretion
disk to explain  the $\Gamma-L_{\rm bol}/L_{\rm Edd}$ anticorrelation,
which is in contrast to  the one observed in luminous AGN \citep[][and
  references therein]{shemmer08apj:gamvsedd}.   Additionally, we would
like to  point out  that very recently,  \citet{xu11:AoxLLAGN} reached
similar results  when considering a  sample of 49 LLAGN,  including 21
LINERs, that  exhibit $L_{\rm bol}/L_{\rm  Edd}<10^{-3}$.  In contrast
to  our work,  where  UV  luminosities are  derived  directly from  UV
observations plus being simultaneous to X-rays, the author derived the
2500~\AA\ luminosities  for their sources by extrapolating  the B band
optical  luminosity, calculated either  directly or  indirectly (using
the   luminosities   of   the   H$\alpha$  and/or   H$\beta$   lines).
Nevertheless, the slope  of the correlation that \citet{xu11:AoxLLAGN}
found (0.163)  is in good agreement,  within the error  bars, with our
result.  \citet{xu11:AoxLLAGN}  was able to  ``roughly'' reproduce the
correlation using the advection dominated accretion flow (ADAF) model.
In ADAFs, the  optical/UV luminosity is the result  of inverse Compton
scatter of  the soft synchrotron photons  by the hot  electrons in the
flow.  The X-ray  photons result from second order  inverse Compton of
soft synchrotron photons by  the hot electrons and from bremsstrahlung
processes.    At  high  accretion   rates  (which   is  conservatively
equivalent  to the  Eddington  ratio), the  inverse Compton  component
dominates  the X-ray  spectrum.  With  decreasing accretion  rate, the
inverse Compton component  becomes softer (due to the  decrease in the
Compton $y$-parameter,  Paper~1), and  X-ray photons result  only from
bremsstrahlung process.   Hence, the  X-ray flux will  decrease faster
than  the  UV flux  with  decreasing  accretion  rate (see  Fig.~5  of
\citealt{xu11:AoxLLAGN}  where the author  plots the  ADAF predictions
for three different acceretion rates of a given BH mass), resulting in
the above seen \alphaox--Eddington~ratio correlation.  This last point
adds to the  long list of evidence supporting  RIAFs as accretion-mode
candidates in LLAGN.

Finally, we have stated in \S~\ref{opt-prop-sec} that we have excluded
two LINERs  from \citet{maoz07MNRAS} sample that  clearly diverge from
the  \alphaox--Eddington-ratio  relation.  These  two  LINERs are  the
faintest  between  all  LINERs  considered  here,  having  the  lowest
\eddratio\ ($<10^{-7}$), but with a somewhat high \alphaox\ ($>-1.1$).
Three possibilities emerge: (1) this  is due to the variability factor
since   the  UV   and   X-ray  fluxes   of  \citet{maoz07MNRAS}   were
non-simultaneous, (2) the UV emission  of these two sources is heavily
internally  absorbed, hence correcting  for internal  extinction would
drive    the    \alphaox\    to    lower   values,    or    (3)    the
\alphaox--Eddington-ratio     relation    breaks    at     very    low
\eddratio\  ($\lesssim10^{-7}$), i.e., during  the quiescent  state of
AGN.  This  would imply that  a different radiative process  is taking
place at such  very low Eddington ratios.  This  latter speculation is
interesting     to     the     fact     that     \citet[][see     also
  \S~\ref{sec-fund-plane-chap5}]{  yuan05apj:jetvsriaf} predicted that
under a  critical \eddratio\ of about $\sim10^{-6}$,  i.e., during the
quiescent state of AGN, the jet emission should dominate over the RIAF
from  radio  to  X-rays.   A  bigger  sample  of  AGN  with  very  low
\eddratio\ would help confirm or refute this hypothesis.

\subsection{Radio emission in LINERs and the role of the jet}
\label{sec-rad-jet}

It is now  believed that radio emission is  a common characteristic of
low  luminosity  AGN  in   general  and  LINERs  specifically  \citep{
  nagar00apj:radioliner,    falcke00ApJ:VLBAliners,    nagar01:radobs,
  nagar02aap, 4278nagar05aap}.  For a  given X-ray or UV emission, the
radio  luminosities  of  these  type  of  objects  compare  well  with
radio-loud  sources, and  exhibit  radio-loudness parameters,  $R_{\rm
  o(UV)}$ and $R_{\rm X}$, that belong to the radio-loud population of
AGN \citep{ho99sed,ho02ApJ:radLLAGN,maoz07MNRAS,eracleous10:linersed}.
Our  SED  analysis   strengthen  this  idea,  as  we   show  that  the
radio-loudness  parameter $R_{\rm  X}$ indicates  that all  of  our 11
LINER~1s  with  5~GHz  detection  could be  considered  as  radio-loud
sources  according   to  the  \citet{terashima03apj:rloud}  criterion,
$\log R_{\rm X}>-4.5$.

The  emission processes of  such a  radio emission  is not  yet firmly
understood,  though more  and more  hints are  pointing towards  a jet
and/or  outflow  synchrotron  origin.   \citet{dimatteo01ApJ:jetliner}
showed that synchrotron emission of the thermal relativistic electrons
in a hot accretion flow (e.g., RIAF) over-estimates the radio emission
of the nuclei of his  sample of four elliptical galaxies.  The authors
assumed an  accretion rate  in the RIAF  equal to the  Bondi accretion
rate calculated from the typical temperatures and densities of the hot
gaseous halos surrounding their  sample nuclei. Even with an accretion
rate  much less  than  the Bondi  rate,  the radio  spectral shape  is
inconsistent with the RIAF  prediction. However, if the accretion flow
produces outflows  and/or jets  in the inner  regions, as  is expected
from RIAF  models \citep{narayan95:jet,blandford99MNRAS:jet}, emitting
synchrotron  non-thermal  emission,   the  radio  spectral  shape  and
luminosities    reconcile     with    the    observations.     \citet{
  ulvestad01ApJ:jetAdaf} came to the same conclusion when studying the
radio  emission  of three  low  luminosity  AGN  observed at  4  radio
wavelengths.  Moreover,  \citet{nagar01:radobs} carried out  a similar
analysis performed on  a bigger sample of 16  low luminosity AGN. They
found that  a jet model is a  better explanation than a  RIAF model of
the flat  radio spectra. In fact,  a large fraction of  LINERs and low
luminosity AGN  have flat spectra in  the radio band  (e.g., between 2
and 20~cm, \citealt{nagar02aap,anderson04ApJ:jetliner}), and some even
exhibit subparsec scale radio jets \citep{falcke00ApJ:VLBAliners}.

\citet{ho02ApJ:radLLAGN}  found a  strong anticorrelation  between the
radio-loudness  parameter $R_{\rm o}$  and the  Eddington ratio  for a
sample of low luminosity and normal luminous AGN spreading over almost
10   orders  of   magnitude  in   Eddington  ratio   space  \citep[see
  also][]{sikora07apj:radloudagn}.     \citet{   panessa07AA:radllagn}
confirmed the same anticorrelation  for their sample of low luminosity
Seyferts when considering  the $R_{\rm X}$ parameter. We  find in this
work that $\log R_{\rm X}$  is highly anticorrelated with \eddratio\ for
our sample  of LINER~1s (Fig.~\ref{rad-loud-edd}, we  will discuss the
case of NGC~315, which is the only outlier compared to the rest of our
sample, later  this section), thus stretching  this anticorrelation to
include  LINER~1s  specifically, and  AGN-powered  LINERs in  general.
This  behavior  is perfectly  in  line  with  the prediction  of  RIAF
structure to produce and  collimate relativistic jets more efficiently
with decreasing Eddington ratio. 

\citet{panessa07AA:radllagn} found a  positive correlation between the
hard X-ray luminosity and the  5~GHz radio luminosity for their sample
of low luminosity Seyfert galaxies  with a slope of $0.97$. To compare
their results to radio galaxies, the authors collected radio and X-ray
luminosities    for     a    sample    of     LLRGs    from    \citet{
  chiaberge05apj:hstliner}   and  \citet{balmaverde06AA:rad3C}.   They
found that LLRGs show, similar  to low luminosity Seyferts, a positive
correlation  with a  slope of  $0.97$  but three  orders of  magnitude
shifted   towards    radio   luminosities   (Fig.~\ref{rad-loud-edd}).
\citet{panessa07AA:radllagn} pointed out the presence of a gap between
the  two  populations clearly  seen  in  $L_{\rm  r}-L_{\rm X}$  space
(Fig.~\ref{rad-loud-edd}).   These putative objects,  as noted  by the
authors, are none other than LINERs.  This is shown in the right panel
of Fig.~\ref{rad-loud-edd} where the majority of our LINER~1s populate
the space  between the  two best fits  of low luminosity  Seyferts and
LLRGs (again, with the exception  of NGC~315, which clearly belongs to
LLRGs). Although our small number of 11 LINER~1s makes it difficult to
perform rigorous statistical analysis,  we find, using a simple linear
regression analysis,  that $\log L_{\rm  X}\propto(0.8\pm0.2)\log L_{\rm
  r}$ (excluding the NGC~315 radio  galaxy would result in $\log L_{\rm
  X}\propto(1.1\pm0.3)\log L_{\rm  r}$), which is  in agreement, within
the   error  bars   with  the   slope  found   for  both   samples  of
\citet{panessa07AA:radllagn}.   Bigger  samples  of  LINERs  would  be
better  suited for  such analysis.   Nonetheless, similar  slopes were
found for a  sample of low-power radio galaxies  with X-ray data taken
from {\sl ROSAT}  \citep{canosa99MNRAS:radGal} and for the radio-quiet
sample  of  \citet[][a  steeper  slope  of 0.5  was  found  for  their
  radio-loud AGN]{brinkmann00AA:radAGN}.  Such correlations could only
mean  that a  coupling between  the radio  emission mechanism  and the
X-ray emission mechanism exists.  Whether both emissions emanates from
the  same  origin (i.e.,  jet),  or  emanates  from different  origins
although highly coupled (i.e.,  jet-RIAF) is a highly debated subject,
which  brings  us to  the  ``fundamental  plane  of BH  activity'',  a
reliable tool  to distinguish  X-ray processes of  different accreting
BHs.

Finally we  would like  to give  our thoughts on  the special  case of
NGC~315. The  LINER~1 NGC~315 is the  only source in our  sample to be
classified as  a radio galaxy (FR~I morphology),  showing an arcsecond
scale radio and X-ray jet. It  also harbors the second most massive BH
in  our sample  with $M_{\rm  BH}>10^9$~M$_{\odot}$.   NGC~315 clearly
does  not follow  the  anticorrelation shared  by  the other  LINER~1s
between  the  radio-loudness, $R_{\rm  X}$,  and  the Eddington  ratio
(Fig.~\ref{rad-loud-edd}).  This behavior  of radio galaxies to occupy
a  parallel,  towards  radio-louder  systems, anticorrelation  in  the
radio-loudness--Eddington-ratio      space     was      noticed     by
\citet{sikora07apj:radloudagn}.   The BH mass  has been  attributed in
many studies as a major factor behind the dichotomy in the strength of
the radio  emission between  radio galaxies and  normal AGN,  with the
most     massive    BHs     resulting     in    radio-loud     sources
\citep[e.g.,][]{laor00ApJ:radBHmass,  dunlop03MNRAS:radqso}.   Indeed,
\citet{broderick11:radAGN}   showed  that   the  two   populations  of
\citet{sikora07apj:radloudagn} almost overlap if  the BH mass is taken
into account and only the  core radio emission is considered. However,
\citet{ho02ApJ:radLLAGN}  noted  that   radio-loud  sources  could  be
present  at  the  center of  AGN  harboring  BHs  of very  low  masses
($10^{6}$~M$_{\odot}$), plus  relatively radio-quiet sources  could be
present in  AGN with  BH masses $>10^9$~M$_{\odot}$  (e.g., NGC~3998).
\citet{sikora07apj:radloudagn} introduced  the BH spin  along with the
BH mass as a driver of the radio-loudness in AGN (as has been proposed
in    the    past,   \citealt{blandford90agn,    wilson95ApJ:radiogal,
  meier99ApJ:jetpro}).  Indeed, FR~I  radio galaxies, to which NGC~315
belongs,    are    believed    to    have   rapidly    spinning    BHs
\citep{wu11ApJ:FRIBHspin}. Consequently, the BH mass, the BH spin, and
the Eddington ratio might  be behind the dichotomy between radio-quiet
and radio-loud AGN.

\subsection{Fundamental plane of LINER~1s, RIAF versus jet}
\label{sec-fund-plane-chap5}

It has been almost 8 years  now since the discovery of the fundamental
plane of BH  activity uniting, X-ray luminosities, radio-luminosities,
and BH  masses of different types  of accreting BHs,  being stellar or
supermassive,     thanks     to     the     pioneering     work     of
\citet{merloni03MNRAS:BHfunplane}.  Since then, numerous articles have
seen the light as a complement or to better constrain the coefficients
of     such    a    fundamental     plane    \citep[][to     name    a
  few]{kording06A&A:funpla,      wang06ApJ:funpla,     li08ApJ:funpla,
  gultekin09ApJ:funpla,     yuan09ApJ:fundplane,    plotkin11:funpla}.
\citet{  merloni03MNRAS:BHfunplane} used a  large sample  of different
classes of accreting  BHs, including XRBs (being in  both low/hard and
high/soft states),  low luminosity  AGN, Seyferts, and  quasars.  They
found that  the coefficients of  their ``fundamental plane''  could be
reproduced  in a  RIAF context  at the  1 sigma  confidence  level.  A
geometrically thin  accretion disk is  excluded at the 3  sigma level,
whereas  a jet  is roughly  consistent with  the data  at the  3 sigma
level.  Mixing  quasars, Seyferts, and  high/soft state XRBs  with low
luminosity AGN and  low/hard state XRBs could be  misleading.  XRBs in
their low/hard state  show a very strong radio  and X-ray correlation,
which breaks when these XRBs are in a high/soft state \citep[e.g.,][]{
  fender99ApJ:HSXRB}.  Moreover,  Seyferts and quasars  are thought to
have  a X-ray  emission mechanism  different than  the one  thought to
exist in low  luminosity AGN (Paper~1).  This could,  in part, explain
the relatively  large scatter around  the coefficients of  the authors
fundamental     plane.     At    the     time    of     the    \citet{
  merloni03MNRAS:BHfunplane}   discovery,  \citet{falcke04A&A:jetonly}
showed  that  his  sample of  Galactic  BH  binaries  in the  low  and
quiescent  state, LINERs,  FR~I  radio galaxies,  and  BL Lac  objects
(i.e., low  Eddington ratio radio  galaxies with the jet  aligned with
our line of sight)  can be unified on a 3D plane,  none other than the
fundamental plane.  The authors  explained the scaling coefficients of
their  relation with  the analytical  solution  of a  jet model.   The
fundamental  plane discovery, in  itself, is  a gigantic  step towards
understanding accretion processes  and emission mechanisms giving rise
to the  radio and  X-ray luminosities. Thus,  refining the  plane with
less scatter around the coefficients is crucial to distinguish between
the different  accretion models in  low accreting BHs,  mainly between
the two most competing ones ``RIAF versus jet''.

\begin{figure*}[!t]
\begin{center}
\includegraphics[angle=0,totalheight=0.20\textheight,width=0.49\textwidth]{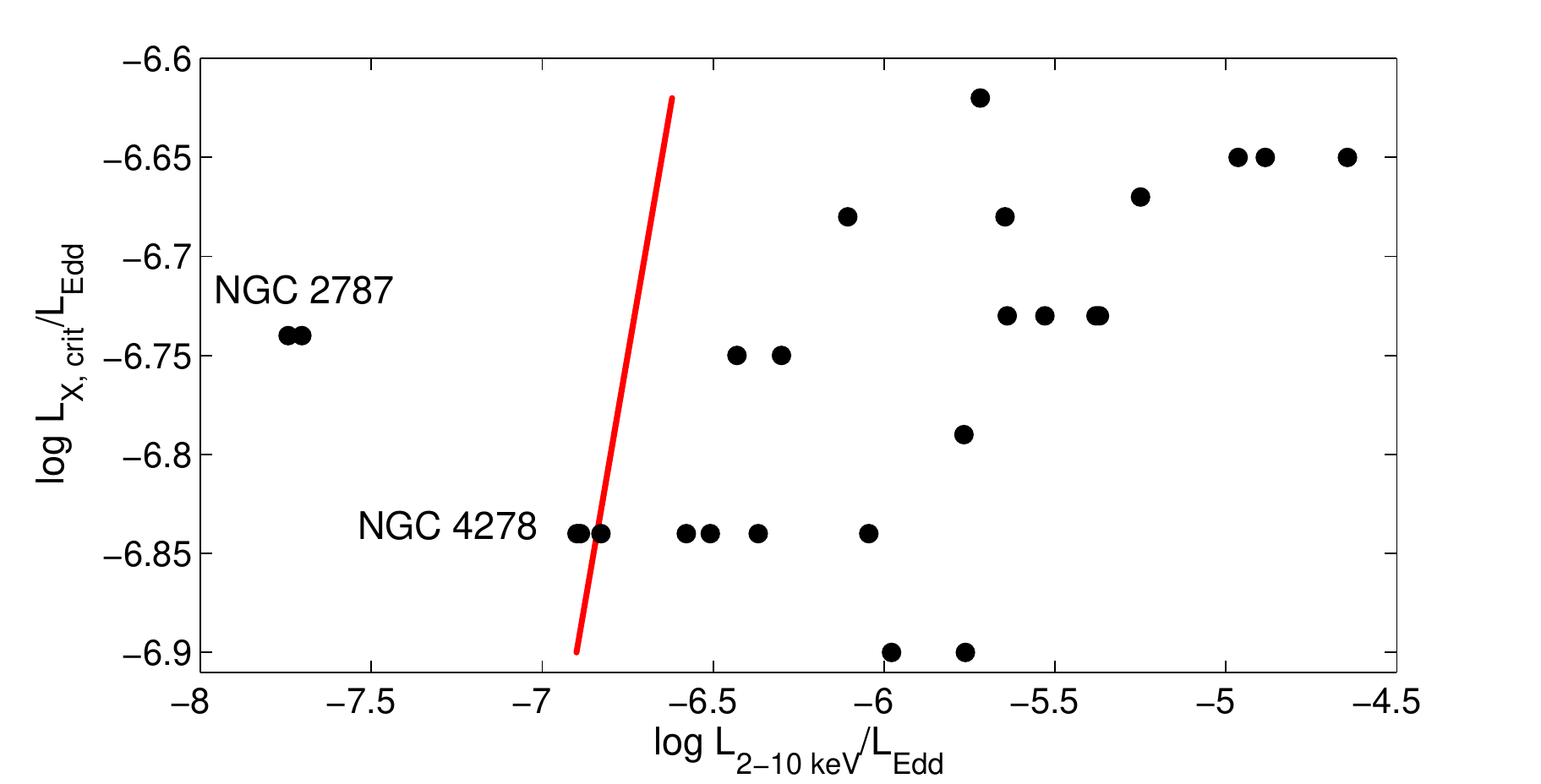}
\includegraphics[angle=0,totalheight=0.20\textheight,width=0.49\textwidth]{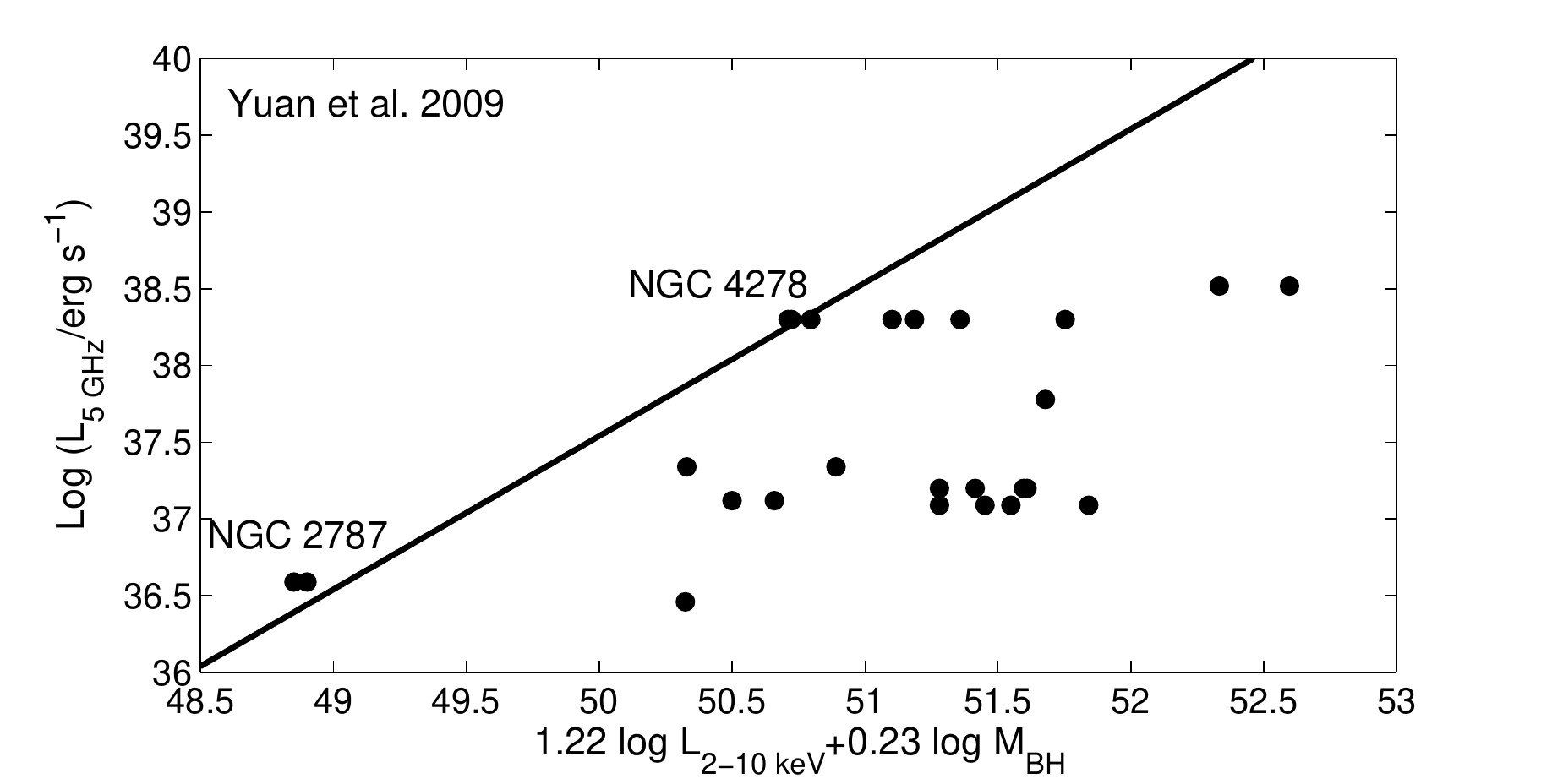}
\caption{{\sl Left panel.} Solid line represents the distribution of the critical X-ray luminosity, $L_{\rm X, crit}/L_{\rm Edd}$, below (above) which the X-ray emission is thought to be jet (RIAF) dominated \citep{yuan05apj:jetvsriaf}. Black dots represent the distribution of \eddratio\ for our sample of 10 LINER~1s (Paper~1). Only NGC~2787 and the three lowest flux NGC~4278 observations have lower \eddratio\ than the critical one. {\sl Right panel.} Our LINER~1s data points superimposed over the \citet{yuan09ApJ:fundplane} fundamental plane (black solid line). Only the observations with \eddratio$<L_{\rm X, crit}/L_{\rm Edd}$ follow the authors fundamental plane, in perfect agreement with their prediction. See text for more details.}
\label{yuan-us-fp}
\end{center}
\end{figure*}

 Our small sample of 10  LINER~1s\footnote{We decide to exclude the FR
   I radio galaxy  NGC~315 from this analysis since  these sources are
   not  appropriate to  test the  jet-only  model, for  the reason  of
   synchrotron  cooling  occurring  at  very low  frequencies  \citep{
     fossati98MNRAS:synchCutoff, kording06A&A:funpla}.}  with detected
 5~GHz cores do  not allow us to perform  a rigorous fundamental plane
 analysis.  Therefore, in order  to infer the X-ray emission mechanism
 and to favor a  RIAF or a jet model over the  other, we relied on two
 different    works:     \citet{yuan09ApJ:fundplane}    and    \citet{
   plotkin11:funpla}.

\citet{yuan09ApJ:fundplane}  studied the fundamental  plane of  22 low
luminosity AGN with X-ray  luminosities below a critical value $L_{\rm
  X, crit}$, defined as:

\begin{equation}
\log\left(\frac{L_{\rm X, crit}}{L_{\rm Edd}}\right)=-5.356-0.17 \log\left(\frac{M}{M_{\odot}}\right),
\label{yuaneq}
\end{equation}

\noindent below (above) which the  X-ray emission should be jet (RIAF)
dominated,  according to  the  \citet{yuan05apj:jetvsriaf} prediction.
\citet{yuan09ApJ:fundplane}  found  that their  sample  of LLAGN  with
X-ray  luminosities  below the  critical  value, hence  jet-dominated,
follows the relation

\begin{equation}
\log L_{\rm   R}=1.22~\log L_{\rm  X}+0.23~\log M_{\rm  BH}-12.46.
\label{eqfunplyuan}
\end{equation}

Recently,     \citet[][see     also    \citealt{kording06A&A:funpla}]{
  plotkin11:funpla} calculated the fundamental plane of a well defined
sample of XRBs in their low/hard  state, low luminosity AGN and BL Lac
objects. They used  a rigorous statistical method based  on a Bayesian
regression  to  calculate  the  errors  on the  coefficients  and  the
dispersion  of their  equation.  Their  fundamental plane  follows the
equation:

\begin{equation}
\log L_{\rm X}=1.45~\log L_{\rm R}-0.88~\log M_{\rm BH}-6.07,
\label{eqplotkin}
\end{equation}

\noindent with a  dispersion $\sigma=0.07$~dex\footnote{Note here that
  \citet{plotkin11:funpla}  choose their  X-ray luminosities  as their
  dependent variable,  the traditional  form of the  fundamental plane
  that  we  choose to  use  here is  easily  recovered  with a  simple
  variable  substitution, this gives  $\log L_{\rm  R}=0.69\log L_{\rm
    X}+0.61\log  M_{\rm  BH}+4.19$}.    The  authors  found  that  the
coefficients of  the fundamental plane  of their sample are  very well
explained  with  a jet  model  \citep[the  scale-invariant jet  model,
][]{heinz03MNRAS:jetADAF}   and   that   a  RIAF   model   \citep[ADAF
  solution,][]{narayan94ApJ:adaf} is excluded at the 3 sigma level.

We  first  plot  in   the  left  panel  of  Fig.~\ref{yuan-us-fp}  the
distribution of  the critical X-ray luminosities (red  solid line, see
also Table~\ref{alphaOX-Ruv-Rx-tab}) of  the different LINER~1s in our
sample, below  (above) which the  X-ray emission should be  jet (RIAF)
dominated, according to  \citet{yuan05apj:jetvsriaf}.  We overplot the
X-ray luminosities  of the different X-ray observations  of our sample
(black dots, see  Paper~1 for the different X-ray  luminosities of our
sample).   The left  panel  of Fig.~\ref{yuan-us-fp}  shows that  only
NGC~2787  and the three  \chandra\ observations  of NGC~4278  with the
lowest  fluxes (obs.   IDs: 7077,  7080, and  7081, see  Paper~1) have
\eddratio\ smaller  than $L_{\rm X,  crit}/L_{\rm Edd}$. In  the right
panel,  we  plot  our  data  points  with  the  fundamental  plane  of
\citet[][black    solid    line,   see    equation~\ref{eqfunplyuan}]{
  yuan09ApJ:fundplane}.  The  results are  perfectly in line  with the
\citet{yuan09ApJ:fundplane}  prediction, in  the sense  that  only the
sources with  \eddratio$<L_{\rm X, crit}/L_{\rm  Edd}$, i.e., NGC~2787
and  the three  NGC~4278 observations  with the  lowest  X-ray fluxes,
follow their  fundamental plane,  with the rest  of the  sources being
inconsistent with it.  According to  the authors work, this would mean
that the  sources in our  sample are consistent with  a RIAF-dominated
X-ray emission process,  with the exception of NGC~2787  and the three
NGC~4278 observations with the lowest  fluxes, where a jet is probably
dominating the X-ray emission.

\begin{figure}[!t]
\begin{center}
\includegraphics[angle=0,totalheight=0.20\textheight,width=0.49\textwidth]{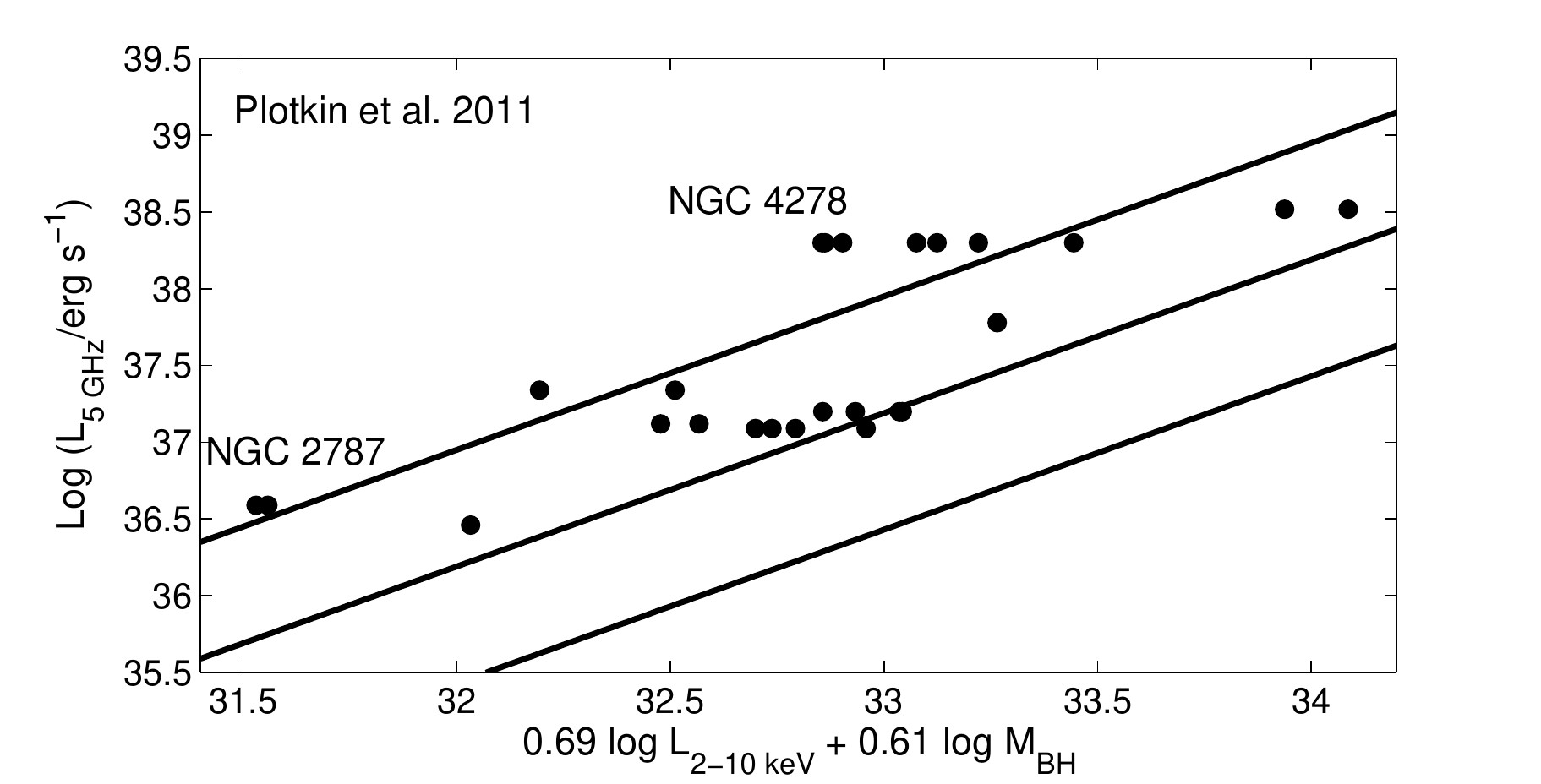}
\caption{Our sample of LINER~1s plotted with the fundamental plane of \citet{plotkin11:funpla}. At the 1$\sigma$ level, our sample is consistent with the authors fundamental plane, with the mild exception of the NGC~4278 observations with the lowest X-ray fluxes, and NGC~2787. See text for more details.}
\label{mer-plot-us-fp}
\end{center}
\end{figure}

In Fig.~\ref{mer-plot-us-fp}, we show  our LINER~1 data points plotted
alongside  the  \citet{plotkin11:funpla}  fundamental plane.   At  the
1$\sigma$ level, our sample follow well the authors fundamental plane,
with the only  exception being NGC~2787 and the  lowest NGC~4278 X-ray
fluxes  (note,  however,  that   these  exceptions  disappear  at  the
2$\sigma$ level).  \citet{plotkin11:funpla} found that the coefficient
of their fundamental plane are  better explained with a jet model than
the  RIAF model (see  above).  Therefore,  according to  these authors
fundamental plane study,  a jet synchrotron process appears  to be the
mechanism dominating the X-ray emission in our sample of LINER~1s.

The results we get looking at two different fundamental plane analysis
came  contradictory with  one pointing  toward a  RIAF-dominated X-ray
emission  process and  the  other pointing  toward  a jet  domination.
Hence,  using the  fundamental plane  by itself  to  distinguish X-ray
processes is very challenging (as said by \citealt{plotkin11:funpla}),
and clearly  both a RIAF  and a jet  are at work with  strong coupling
between them.   Nevertheless, it would  be of great interest  to study
the  fundamental plane  of a  big-enough, pure  sample  of AGN-powered
LINERs,  preferentially covering  a large  range in  BH  masses (e.g.,
three  orders of  magnitude), comparing  the  correlation coefficients
with the  ones expected from different  radiative processes.  Finally,
before making any firm conclusion regarding the issue, one should look
into   all  the   different  observational   pieces   together  (e.g.,
$\Gamma$--Eddington-ratio, Paper~1, and the \alphaox--Eddington-ratio,
this   work,  relations),   then   consider  explaining   them  in   a
self-consistent manner.

\section{Conclusion}

We have  studied the multiwavelength characteristics of  the sample of
LINERs  showing  a  definite  detection of  broad  H$\alpha$  emission
introduced in  Paper~1 (LINER~1s, \citealt{ho97apjs:broadHal}).  Since
the  nuclear emission from  LINERs could  easily be  contaminated from
off-nuclear  sources,  high spatial  resolution  are  required at  all
wavelengths.   Moreover,  AGN-powered LINERs  show  a  high degree  of
variability,  especially   in  the   UV  and  X-ray   bands  (Paper~1,
\citealt{maoz05apj:linervarUV}),  hence simultaneous  observations are
crucial.  Keeping  these two  points  in  mind,  we collect  from  the
literature  VLA subarcsecond  or  VLBI milliarcsecond  radio data  and
subarcsecond \hst\  optical data. UV  fluxes were derived from  the OM
instrument onboard  \xmm, and are  simultaneous with the  X-ray fluxes
derived from the EPIC instrument.

We build the  SED of six sources in our  sample with {\it simultaneous
  UV and X-ray measurements}.  We compare these SEDs, as well as their
geometric  mean  SED,  to the  SED  of  a  sample of  radio-quiet  and
radio-loud quasars. At a given  X-ray luminosity we find that, (1) our
LINER~1s have radio luminosities comparable to the radio luminosity of
radio-loud quasars,  (2) the mean  optical and UV luminosities  are on
average 5 to 10 times,  respectively, smaller than in the quasar case,
and (3) the  X-ray spectral shape is similar to  the spectral shape of
radio-quiet quasars.

We calculate the UV to X-ray flux ratio, $\alpha_{\rm ox}$, of the six
LINER~1s with simultaneous UV  and X-ray fluxes and found $\alpha_{\rm
  ox}=-1.17\pm0.02$ with  a dispersion $\sigma=0.01$,  indicative of a
weak UV  excess relative  to X-rays.  We  complement our  results with
$\alpha_{\rm   ox}$  values   of  a   sample  of   AGN-powered  LINERs
\citep{maoz07MNRAS,pianmnras10}.  We find that \alphaox\ is positively
correlated    to   the    Eddington   ratio    (considering   negative
\alphaox\  values),  in  contrast  to  the  relation  established  for
luminous   AGN  \citep[e.g.,   ][]{lusso10A&A:oxSEY}.   This   may  be
indicative of  different emission processes producing the  UV to X-ray
spectral shape between AGN and LINERs.

Using high  resolution radio fluxes  at 5~GHz, we calculate  the radio
loudness  parameter, $R_{\rm X}=\nu  L_{\nu}(5~GHz)/L_{\rm 2-10~keV}$,
for  11  of the  13  LINER~1s  having  radio core-emission  detection.
According to the \citet{terashima03apj:rloud} criterion, all of the 11
LINER~1s  are considered  radio-loud having  $\log R_{\rm  X}>-4.5$. In
fact 10  out of 11 LINER~1s  have $\log R_{\rm  X}>-3.5$.  We establish
for the first  time for LINER~1s, that this  radio loudness parameter,
$R_{\rm  X}$,  is  strongly  anticorrelated to  the  Eddington  ratio,
confirming  previous  studies  on  LLAGN  and  luminous  AGN  sources.
Moreover, we find a  positive correlation between the radio luminosity
and  the  X-ray  luminosity  which  places AGN-powered  LINERs,  on  a
radio-power  scale,  right between  low  luminosity  Seyferts and  low
luminosity radio galaxies.

Finally, we  attempted to  infer the X-ray  emission mechanism  in our
sample of LINER~1s with the help of two different ``fundamental planes
of BH  activity''.  The  results were contradictory.   The fundamental
plane of  \citet{yuan09ApJ:fundplane} pointed toward  a RIAF-dominated
origin  for the  X-ray emission  in our  sample.  On  the  other hand,
\citet{plotkin11:funpla} fundamental  plane indicates that  our sample
of LINER~1s may have a jet dominated X-ray emission.

\acknowledgements This research has made use of observations made with
the  NASA/ESA Hubble  Space Telescope,  and obtained  from  the Hubble
Legacy Archive,  which is a collaboration between  the Space Telescope
Science   Institute  (STScI/NASA),   the   Space  Telescope   European
Coordinating  Facility (ST-ECF/ESA)  and the  Canadian  Astronomy Data
Centre  (CADC/NRC/CSA). This research  has made  use of  the NASA/IPAC
Extragalactic Database  (NED) which is operated by  the Jet Propulsion
Laboratory,  California Institute of  Technology, under  contract with
the National Aeronautics and  Space Administration.  This research has
made use of the data obtained  from the {\sl Chandra Data Archive} and
the {\sl  Chandra Source Catalog},  and software provided by  the {\sl
  Chandra  X-ray Center} (CXC)  in the  application packages  CIAO and
Chips.  This work  is based on observations with  \xmm, an ESA science
mission  with instruments  and  contributions directly  funded by  ESA
Member States  and the USA (NASA).  This research has made  use of the
SIMBAD database, operated at  CDS, Strasbourg, France. G.Y. would like
to thank Yadi Xu for enlightening discussions on ADAF models.


\end{document}